\documentclass[fleqn,usenatbib]{mnras}
    
    \usepackage{newtxtext,newtxmath}
    
    \usepackage[T1]{fontenc}

    \DeclareRobustCommand{\VAN}[3]{#2}
    \let\VANthebibliography\thebibliography
    \def\thebibliography{\DeclareRobustCommand{\VAN}[3]{##3}\VANthebibliography}

    \usepackage{graphicx}	
    \usepackage{amsmath}	
    \usepackage{makecell}
    
    \usepackage{booktabs}
    \usepackage{hyperref}   
    \usepackage{dcolumn}
    \usepackage{makecell}
    \usepackage{tabularx}
    \usepackage{subcaption}
    
    \usepackage{rotating}
    \usepackage{cancel}

    \title{Effect of spin in binary neutron star mergers}

    \author[Beyhan Karakas, Rahime Matur, and Maximilian Ruffert]{
        Beyhan Karakas,$^{1}$\thanks{E-mail: \href{mailto:beyhannkarakas@gmail.com}{beyhannkarakas@gmail.com}; Fellow of the Royal Astronomical Society}
        Rahime Matur,$^{2}$\thanks{E-mail: \href{mailto:r.matur@soton.ac.uk}{r.matur@soton.ac.uk}}
        Maximilian Ruffert$^{3}$
        \\
        $^{1}$beyhannkarakas@gmail.com\\
        $^{2}$Mathematical Sciences and STAG Research Centre, University of Southampton, Southampton SO17 1BJ, UK\\
        $^{3}$School of Mathematics and Maxwell Institute, University of Edinburgh, Edinburgh EH9 3FD, UK\\
    }
    
    \date{Accepted XXX. Received YYY; in original form ZZZ}
    
    \pubyear{2025}
    
\begin{document}
    \label{firstpage}
    \pagerange{\pageref{firstpage}--\pageref{lastpage}}
     \maketitle
    
\begin{abstract}
We investigate the effect of spin on equal and unequal mass binary neutron star mergers using finite-temperature, composition-dependent Steiner-Fischer-Hempel equation of state with parameter set ``o'' (SFHo), via 3+1 general relativistic hydrodynamics simulations which take into account neutrino emission and absorption. Equal mass, irrotational cases that have a mass of $M_{1,2}$ =$1.27M_{\odot}$, result in a long-lived neutron star, while $1.52$ and $2.05M_{\odot}$ cases lead to a prompt collapse to a black hole. For all cases, we analyse the effect of initial spin on dynamics, on the structure of the final remnant, its spin evolution, the amount and composition of the ejected matter, gravitational waves, neutrino energies {and luminosities}, and disc masses. We show that in equal mass binary neutron star mergers, the ejected mass could reach $\sim0.06M_{\odot}$ for highly aligned-spins ($\chi=0.67$). The black hole which results from such a highly spinning, high-mass binary neutron star merger reaches a dimensionless spin of $0.92$; this is the highest spin reached in binary neutron star mergers, to date.
\end{abstract}


\section{Introduction}\label{introduction}
    
The detection of the gravitational wave (GW) event \texttt{GW170817}~\citep{abbott_gw170817_2017} by the Advanced LIGO~\citep{the_ligo_scientific_collaboration_advanced_2015} and Advanced Virgo~\citep{acernese_advanced_2015} detectors corresponded to the inspiral phase of a binary neutron star (BNS) merger. Its electromagnetic (EM) counterpart, \texttt{EM170817}~\citep{doi:science_em170817}, which included the short Gamma-Ray Burst (sGRB) \texttt{GRB170817A}, provided the first direct confirmation of the long-anticipated association between BNS mergers and sGRBs~\citep{eichler_nucleosynthesis_1989, piran_implications_1992, ruffert_coalescing_1995, ruffert_coalescing_1997, ruffert_coalescing_2001, Abbott_2017_multi,abbott_gravitational_2017, 2017_sgrb,dietrich_science}. Beyond this milestone, \texttt{GW170817} offered pivotal insights into neutron star physics. By assuming low spin priors and using observed neutron star masses, it constrained neutron star spins~\citep{abbott_properties_2019} and the equation of state (EoS), ruling out both very soft and very stiff EoSs. Building on these EoS constraints, the maximum TOV mass is estimated to lie within the range $(1.97M_{\odot}<M_{\mathrm{max}}<2.17M_{\odot})$~\citep{metzger_gw17087constraint, the_ligo_scientific_collaboration_and_the_virgo_collaboration_gw170817_2018, Radice_gw170817,abbott_properties_2019}. Additionally, \texttt{GW170817} strengthened evidence for BNS mergers as sites of r-process nucleosynthesis and kilonovae~\citep{kasen_origin_2017, smartt_kilonova_2017,pian_spectroscopic_2017, troja_x-ray_2017,hallinan_radio_2017}, provided a strong test of general relativity in the strong gravity regime~\citep{abbott_tests_2019} and placed constraints on alternative theories of gravity~\citep{sakstein_implications_2017, baker_strong_2017}. 
    
The current detectors are sensitive mainly to the inspiral phase of BNS mergers, enabling the measurement of three key parameters: chirp mass ${\mathcal{M}}$~\citep{kafka_physics_1988, finn_observing_1993, cutler_last_1993, cutler_gravitational_1994}, effective spin 
$\chi = (M_1 \chi_1 \cos\theta_1 + M_2 \chi_2 \cos\theta_2)/(M_1 + M_2)$, where $\chi_{1,2}$ are the dimensionless spin magnitudes, $\theta_{1,2}$ are the angles between the spin vectors and the orbital angular momentum, and $M_{1,2}$ are the masses of each star, and tidal deformability parameter of the binary $\Lambda$,~\citep{hinderer_tidal_2008, flanagan_constraining_2008, hinderer_tidal_2010, read_matter_2013, abbott_gravitational_2017}. For \texttt{GW170817}, assuming high spin priors for the components aligned with the orbital angular momentum, $(\left|\chi\right| \leq 0.89)$, these parameters were measured as ${\mathcal{M}}=1.188^{+0.004}_{-0.002}M_{\odot}$, $\chi \in (0.0, 0.09)$, and $\Lambda<800$ (at $90$ per cent credible interval)~\citep{abbott_gw170817_2017}. Among these, tidal deformability is particularly important for extracting EoS information from gravitational waves. However, a recent study by~\citet{RM_problematic} quantitatively demonstrated that this calculation can be affected by artificial heating of the stars, which introduces significant systematics. Unlike \texttt{GW170817}, another BNS merger, \texttt{GW190425}~\citep{abbott_gw190425_2020}, was observed solely via gravitational waves. By assuming high spin priors, the effective spin for \texttt{GW190425} was measured as $\chi=0.058^{+0.11}_{-0.05}$.

To accurately interpret gravitational wave observations in BNS mergers, it is essential to perform numerical simulations with well controlled and systematically varied spin configurations, starting from consistent initial data. Theoretical efforts to create spinning BNS initial data (ID) began with~\citet{marronetti_relativistic_2003}, and were further explored in studies by~\citet{tichy_black_2006, tichy_new_2009, tichy_long_2009, tichy_initial_2011,tichy_constructing_2012, kastaun_black_2013, tsatsin_initial_2013, tacik_binary_2015, tichy_constructing_2019, papenfort_new_2021, elliptica_bns}. While these studies focused on constructing spinning ID, state-of-the-art BNS merger simulations now address increasingly complex aspects of these systems, including turbulence modelling, neutrino transport and computational efficiency. Key advancements include subgrid-scale turbulence modelling, which serves as an alternative to ultra-high resolution general relativistic magnetohydrodynamics simulations~\citep{Radice_turbulence2017, Radice_viscousejecta2018, Radice_turbulence2020}, see also~\citet{Radice_Ian_Review2024} for a recent review on the impact of turbulence in BNS mergers.  Additional advancements include improved vacuum treatment~\citep{vacuum_2020}, using discontinuous Galerkin method~\citep{tichy_dg2023, Foucart_dg2024}, and the combination of fixed mesh-refinement with smoothed particle hydrodynamics~\citep{rosswog_diener_frontiers, rosswog_2024}. Furthermore, improved neutrino transport models have been developed~\citep{Foucart_neutrino2021, Radice_M1_2022, Foucart_neutrino2024}, with their significance highlighted in a recent review~\citep{Foucart_review2023}. Finally, GPU-based simulations have been reported to demonstrate an order-of-magnitude speed-up over CPU-based simulations of these systems~\citep{Fields_gpu2024}.

Building on advancements in BNS merger simulations, several studies have investigated the impact of spin. Among these,~\citet{kastaun_black_2013}, was the first to investigate how the initial spin affects the maximum spin of the final black hole (BH) in equal mass models that promptly collapse to a BH. They reported an upper limit of $\chi = 0.88\pm0.018$ and identified, for the first time, the orbital hang-up effect, a repulsive spin-orbit interaction observed in systems with spins aligned with the orbital angular momentum. This phenomenon had previously been reported in binary black hole (BBH),~\citep{campanelli_spinning-black-hole_2006}, and black hole-neutron star mergers~\citep{2009PhRvD..79d4024E}. Subsequent investigations by~\citet{bernuzzi_mergers_2014, Dietrich_2017_rotation} confirmed this finding. Notably,~\citet{bernuzzi_mergers_2014} extended these investigations to include anti-aligned spin BNS merger simulations. Their results demonstrated that anti-aligned spin models merge faster, a phenomenon called 'speed-up', as compared to the irrotational model. This behaviour was consistent with earlier findings in BBH mergers~\citep{campanelli_spinning-black-hole_2006}. They also reported a shift in the peak frequency of the fundamental mode $f_{2}$, corresponding to the $(l,m) = (2,2)$ mode, towards lower frequencies for aligned spins of $\chi=0.05$.

While earlier studies used simple Gamma-Law EoSs,~\citet{kastaun_properties_2015} was the first to employ microphysical EoSs to investigate spinning BNS mergers. Their work examined equal mass spinning (aligned) and irrotational models, as well as unequal mass irrotational models that primarily form long-lived neutron stars. They reported that the rotation profile of the remnant differed from predictions based on a single, differentially rotating neutron star, with the remnant core rotating more slowly than the envelope.~\citet{dietrich_binary_2015} presented the first precessing BNS merger simulation, along with the most asymmetric mass ratio, $q = 2.06$, where $q=M_1/M_2 \geq 1$ is defined as the ratio of the mass of the more massive star ($M_1$) to that of the less massive star ($M_2$). They observed modulation in the $(l,m)=(2,1)$ mode of gravitational waves due to precession and reported a shift in $f_{2}$, similar to the findings of~\citet{bernuzzi_mergers_2014}. Similarly,~\citet{tacik_binary_2015} investigated spin effects on orbital dynamics, including aligned, anti-aligned and misaligned spins. They found that although spin direction changes during the late inspiral, its magnitude remains conserved.
    
~\citet{Dietrich_2017_rotation} extended these investigations to equal and unequal mass binaries, $q \leq 1.5$ with aligned, anti-aligned or just one component spins. Their results showed that spin alignment (anti-alignment) with orbital angular momentum increases (decreases) mass ejection compared to irrotational models. They also reported the $f_{2}$ mode frequency shift previously identified in~\citet{bernuzzi_mergers_2014, dietrich_binary_2015}. In equal mass precessing BNS mergers,~\citet{Dietrich_precessing2018} found that spin precession does not influence post-merger GW frequencies and mass ejection.
    
~\citet{East_2019} explored equal mass spinning BNS mergers with spins aligned and anti-aligned with the orbital angular momentum, considering a maximum spin of $\chi=0.33$. They reported that the dependence of the one-arm, $m=1$, instability on spin is weak~\citep{paschalidis_one-arm_2015, east_relativistic_2016, radice_one-armed_2016}. The frequency of the $m=1$ mode, $f_{1}$, which corresponds to $(l,m) = (2,1)$ mode was observed to shift to lower values for aligned spins. Anti-aligned spin models produce more massive ejecta, while aligned spins reduce ejecta mass up to $\chi=0.17$, after which it increases, equaling the ejecta mass of the irrotational model at $\chi=0.3$.
    
~\citet{Chaurasia_Dietrich2020} recently investigated the influence of spin orientation on GW and mass ejection, showing that for equal mass models with the maximum effective spin magnitude of $0.096$ in aligned and anti-aligned spins, the lifetime of the remnant NS depends primarily on the effective spin magnitude rather than the spin orientation. They reported anisotropic ejecta distribution in precessing mergers.~\citet{Dudi_Dietrich2022} studied high spin models, $-0.28\leq\chi\leq0.58$, finding that aligned spins produce more ejecta and disc mass than anti-aligned spins, with disc wind ejecta showing spin dependence. ~\citet{papenfort_extreme} analyzed single-spin aligned models ($\chi_1 = 0.30,\,0.40,\,0.60$) with mass ratios $1 \leq q \leq 1.67$, using the TNTYST and BHB$\Lambda\Phi$ EOSs. They found that higher mass ratio systems yield longer-lived remnants than the equal-mass binaries, and that disc and ejecta masses increase with both mass ratio and spin. The largest amount of dynamical ejecta occurred for $\chi_1 = 0.60$ with $q = 1$. Similarly, focusing on a single-spin aligned model, $\chi_{1}=0.5$, ~\citet{rosswog_2024} investigated equal mass BNS mergers and found that spinning models result in less violent mergers and significantly brighter kilonovae as compared to irrotational models. They also explored the dependence of $f_{2}$ on the EoS, showing that the softest EoS (SLy) exhibited the highest frequency shift, while the stiffest EoS (MS1b) showed no detectable shift. Moreover, ~\citet{federico_dietrich_prompt} investigated the impact of spin on black hole formation using piecewise-polytropic SLy and H4 EoSs, considering both equal and unequal mass binaries ($q=1.38, \ 1.63$), with aligned (up to $\chi = 0.2$), anti-aligned (maximum $\chi = -0.1$), and irrotational models. They found that in cases undergoing prompt collapse to a BH, spin increases the lifetime of the remnant neutron star before collapse, and tends to enhance (suppress) the disc mass for aligned (anti-aligned) spin models.

Although our understanding of BNS mergers has progressed from early efforts to investigate BH formation during the inspiral phase~\citep{wilson_instabilities_1995} to realistic simulations of spinning, magnetized mergers with detailed neutrino emission and absorption, current simulations still have simplifications that remain challenging to overcome.
    
In this study, we extend current spinning BNS merger studies to highly aligned, anti-aligned and mixed spin models. We focus on the effect of varying spin on mergers of equal and unequal mass BNS in quasi-circular orbits. The aim is to perform state-of-the-art simulations of various spin configurations in BNS mergers and investigate whether the spin parameter can be constrained using gravitational waves and possible EM counterparts. The centre-to-centre separation is primarily $40$km, except for three test models, where it is $60$km. We use the finite-temperature, composition dependent Steiner-Fischer-Hempel EoS with parameter set ``o'' (SFHo)~\citep{steiner_core-collapse_2013} and account for neutrino emission and absorption. We consider three total mass configurations, where the equal mass irrotational models results in long-lived NS, or prompt BH formation. 
This paper is organized as follows: Sect.\ref{numerical_setup} outlines the initial data and the numerical methods; Sect.\ref{Analysis} details the analysis methods; the results for models with $M_{\mathrm{tot}}=2.55\,M_{\odot}$ are presented in Sect.\ref{Results}, while results for models with $M_{\mathrm{tot}}=3.05$ and $4.10\, M_{\odot}$ are presented in Sect.\ref{results_other}; the discussion and conclusion in Sect.\ref{conclusion}. Geometrized units, $(G=c=1)$, are used unless otherwise specified.

\section{Numerical Setup}\label{numerical_setup}
    
The ID for all BNS models is created using the Fuka branch of the Kadath library~\citep{kadath_paper, papenfort_new_2021}. Table~\ref{initial-data} summarizes the ID parameters and the corresponding merger times.
    
\begin{table*}
\caption{\label{initial-data}The initial data (ID) parameters and merger times. $M_{\mathrm{tot}}^{\infty} / M_{\mathrm{tot}}^{\mathrm{loc}}$ ($M_{\odot}$) represents the ratio of the total ADM mass computed in isolation to the total ADM mass at the initial separation. $M_{1,2}$ and $M^b_{1,2}$ correspond to the ADM mass in isolation, and the baryonic mass of each star, respectively. $J_{0}$, $f_{0}$ and $\chi_{1,2}$ represent the total angular momentum, the initial orbital frequency of the binary and the dimensionless spin parameter of each star. The quantity $t-t_{\mathrm{merger}}$ shows the difference in inspiral time relative to the reference models. The respective reference models are marked with an asterisk (*) for clarity. Negative (positive) values indicate shorter (longer) inspiral times compared to the reference models.}
\small
\begin{tabular}{@{}lcccccccccccc@{}}
    \hline
    Model & $M_{\mathrm{tot}}^{\infty} / M_{\mathrm{tot}}^{\mathrm{loc}}$ ($M_{\odot}$)  & $M_{1}$ ($M_{\odot}$) & $M_{2}$ ($M_{\odot}$) & $M^b_{1}$ ($M_{\odot}$) & $M^b_{2}$ ($M_{\odot}$) & $J_{0}$ ($M_{\odot}^{2}$) & $f_{0}$ (Hz) & $\chi_{1}$ & $\chi_{2}$ & $t-t_{\mathrm{merger}}$(ms) &  \\
    \hline
    $M255_{00}$&2.5456 / 2.5194 &  1.2728 & 1.2728 & 1.3902 & 1.3902& 6.33425 & 324 &  0.00 &  0.00 & $10.35 (10.11)$ & * \\
    $M255_{\downarrow_{0.4}0}$&2.5456 / 2.5273 &  1.2728 & 1.2728 &1.3902 &1.3902 &5.81142 & 326 & -0.40 &  0.00& $-5.61 (-5.40)$  \\
    $M255_{\uparrow^{0.4} 0}$&2.5456 / 2.5271 &  1.2728 & 1.2728 &1.3902 &1.3902& 6.99839 & 326 & 0.40 &  0.00& $+1.48 (+1.39)$  \\
    $M255_{\downarrow^{0.4} \downarrow^{0.4}}$&2.5456 / 2.5280 &  1.2728 & 1.2728 & 1.3853& 1.3853& 5.21892 & 326 & -0.40 & -0.40& $-6.84 (-6.61)$  \\
    $M255_{\downarrow^{0.4} \uparrow^{0.4}}$&2.5456 / 2.5347 & 1.2728 & 1.2728 &1.3902 &1.3902 & 6.43015 & 326 & -0.40 & 0.40& $-4.85 (-4.69)$    \\
    $M255_{\uparrow^{0.4} \uparrow^{0.4}}$&2.5456 / 2.5272 &  1.2728 & 1.2728 &1.3853 &1.3853 & 7.59055 & 326 & 0.40 & 0.40& $+2.55 (+2.34)$   \\
    $M255_{\downarrow^{0.65} \downarrow^{0.65}}$&2.5456 / 2.5220 &  1.2728 & 1.2728 &1.3701 &1.3701 & 4.46266 & 326 & -0.65 & -0.65& $-5.42 (-5.25)$  \\
    $M255_{\downarrow^{0.65} \uparrow^{0.65}}$&2.5456 / 2.5130 &  1.2728 & 1.2728 &1.3701&1.3701 & 6.38663 & 326 & -0.65 & 0.65& $-2.44 (-2.35)$ \\ 
    $M255_{\uparrow^{0.67} \uparrow^{0.67}}$&2.5456 / 2.4932 &  1.2728 & 1.2728 &1.3695 &1.3695 & 8.39040 & 326 & 0.67 & 0.67& $-0.04 (-0.77)$  \\
    $M305_{{0}{0}}$&3.0500 / 3.0161 &  1.5250 & 1.5250 &1.7028 &1.7028& 8.66939 & 349 &  0.00  & 0.00 & $7.60$ & * \\
    $M305_{\downarrow^{0.4} \downarrow^{0.4}}$&3.0500 / 3.0181 &  1.5250 & 1.5250 &1.6883 &1.6883& 7.00962 & 349 &  -0.40  & -0.40 & $-4.33$  \\
    $M305_{\uparrow^{0.4} \uparrow^{0.4}}$ &3.0500 / 3.0159 &  1.5250 & 1.5250 &1.6883&1.6883 & 10.34024 & 349 & 0.40 & 0.40& $+1.92$    \\
    $M305_{\uparrow^{0.67} \uparrow^{0.67}}$ &3.0500 / 3.0156 &  1.5250 & 1.5250 &1.6687&1.6687 & 11.48328 & 349 & 0.67 & 0.67& $+0.09$    \\
    $M305q205_{{0}{0}}$&3.0500 / 3.0201 &  2.0500 & 1.0000 &2.4215 &1.0670 & 7.64993 & 348 &  0.00   & 0.00& $-0.73$    \\
    $M305q205_{\uparrow^{0.6} {0}}$&3.0500 / 3.0203 &  2.0500 & 1.0000 &2.3391 &1.0670 & 9.95296 & 348 & 0.60 & 0.00& $+0.68$  \\
    $M410_{{0} {0}}$&4.1000 / 4.0425 &  2.0500 & 2.0500 &2.4215 &2.4215 & 14.43615& 385 & 0.00  &  0.00& $4.52$ & * \\
    $M410_{\downarrow^{0.65} \uparrow^{0.65}}$&4.1000 / 4.0445 &  2.0500 & 2.0500 &2.3268 &2.3268 & 14.40156& 385   & -0.65 & 0.65& $-0.82$    \\
    $M410_{\downarrow^{0.65} \downarrow^{0.65}}$&4.1000 / 4.0418 &  2.0500 & 2.0500 &2.3268&2.3268 & 9.71222 & 385 & -0.65 & -0.65& $-1.02$    \\
    $M410_{\uparrow^{0.67} \uparrow^{0.67}}$&4.1000 / 4.0458 &  2.0500 & 2.0500 &2.3237 &2.3237 & 19.26407 & 385 & 0.67 & 0.67& $+1.99$  \\
    \hline
    \end{tabular}
    \end{table*}
    GRHD evolution is performed using \texttt{WhiskyTHC}~\citep{radice_thc_2012, radice_high-order_2014, radice_beyond_2014, 2015ASPC..498..121R}, a finite difference/finite volume High-Resolution Shock-Capturing (HRSC) code that implements the Valencia formulation of the general relativistic hydrodynamics equations~\citep{banyuls_numerical_1997}. Built on the \texttt{Cactus} framework~\citep{allen_cactus_1999, goodale_cactus_2003, Cactuscode:web, Cactusprize:web}, \texttt{WhiskyTHC} employs the \texttt{Carpet} adaptive mesh refinement (AMR) driver~\citep{schnetter_evolutions_2004}. Neutron stars are modeled as perfect fluids, with the energy-momentum tensor given as~\citep{radice_thc_2012, radice_high-order_2014, radice_beyond_2014,2015ASPC..498..121R}:
    \begin{equation}
    \centering
    T^{ab}_{\mathrm{h}} = \rho hu^{a}u^{b} + pg^{ab}
    \label{energy-momentum-eq}
    \end{equation}
where $\rho$ is the rest-mass density, $h = 1 + \epsilon + {p}/{\rho}$ is the specific enthalpy, with $\epsilon$ the specific internal energy, $p$ is the pressure, $u^{a}$ is the fluid 4-velocity, $g^{ab}$ is the metric tensor and $T^{ab}_{\mathrm{h}}$ is the energy-momentum tensor for pure hydrodynamics. For our systems, the atmosphere density and temperature are set to $\rho=6.176\times10^{3}\, \mathrm{g cm}^{-3}$ and $T=0.02$MeV. The conservation of total energy-momentum, including neutrinos, is given by $\nabla_{b}T^{ab}_{\mathrm{rad}} = Qu^{b}$, where $Q$ is the net energy deposition rate due to the absorption and emission of the neutrinos~\citep{radice_binary_2018}. We use the finite-volume HRSC method which employs a $5^{th}$ order monotonicity-preserving scheme (MP5)~\citep{suresh_accurate_1997} for reconstruction and the Harten-Lax-van Leer-Einfeldt (HLLE) Riemann solver~\citep{einfeldt_godunov-type_1988} for flux calculation. Neutrinos are included using M0+Leakage~\citep{radice_dynamical_2016}, which tracks electron neutrinos, $\nu_{e}$, anti-electron neutrinos, ${\overline\nu}_{e}$ and heavy lepton neutrinos, $\nu_{x}$; the latter denotes the collective group of the muon and tau neutrino \& anti-neutrinos. The emission from the optically thick regions is computed via a gray leakage scheme, while the transport and radial propagation in optically thin regions is handled by the M0 scheme over a spherical grid(see~\citet{radice_dynamical_2016} for the detailed explanation of the neutrino treatment). The average energy and luminosity of free-streaming neutrinos is calculated on a uniform spherical grid with radius $\sim756$km and size $(r, \theta, \phi)=(3096, 32, 64)$, using $2048$ rays. We use the finite-temperature, composition-dependent SFHo EoS~\citep{steiner_core-collapse_2013}, which is considered soft as it yields for a typical neutron star with mass $M = 1.4 M_{\odot}$ a radius of $\sim 11.9$km. The EoS is fully hadronic, and is available on~\citet{stellarcollapse},~\citet{2010CQGra..27k4103O}.
    
Spacetime evolution is performed with \texttt{CTGamma}~\citep{pollney_high_2011}, which is based on the publicly available software platform \texttt{Einstein Toolkit}~\citep{loffler_einstein_2012, zilhao_introduction_2013, Roland_et2022,EinsteinToolkit:web}. We use the constraint damping Z4c formulation of the Einstein field equations by~\citet{bernuzzi_constraint_2010}, which is used within \texttt{CTGamma}, with moving puncture gauge conditions. The comparison of Z4c and Baumgarte-Shapiro-Shibata-Nakamura-Oohara-Kojima (BSSNOK) formulations~\citep{nakamura_general_1987, shibata_evolution_1995, baumgarte_numerical_1998} shows that the former has a substantially lower constraint violation, more accurate gravitational wave phase and amplitude (\citet{weyhausen_constraint_2012, hilditch_compact_2013} and references therein). The coupling between spacetime and GRHD variables is handled by the \texttt{Method of Lines (MoL)}. We use the strong stability preserving (SSP) $3^{rd}$ order \texttt{Runge-Kutta} method~\citep{gottlieb_high_2009, Radice_turbulence2020} for time integration. The timestep factor is chosen according to the \texttt{Courant-Friedrichs-Lewy (CFL)} condition to be $0.15$.
    
We use a cell-centred grid structure extending to $\approx2835$km in all three directions, with reflection symmetry applied along the $z$ axis to reduce the computational cost. We use $7$ refinement levels, the finest grid of which has a resolution of $\approx308$m which is denoted as LR, low resolution, for all cases, except for $M_{\mathrm{tot}}=2.55M_{\odot}$ models which are also simulated with a resolution of $\approx222$m, denoted as HR, high resolution. The \texttt{"Sophie Kowalevski"} release of the \texttt{Einstein Toolkit}~\citep{Roland_et2022} is used. 
    
\section{Analysis Methods}\label{Analysis}
    
We perform post-processing using \texttt{PostCactus}~\citep{kastaun_pycactus_2021} and \texttt{Scidata}~\citep{scidata}. Unless otherwise stated, the presented plots and figures are based on LR simulations.
    
\textit{Naming and resolution :} Models are named based on total mass, mass ratio for unequal mass models, and spins. For example, $M255_{00}$ refers to an equal mass, irrotational model with $M_{\mathrm{tot}}=2.55M_{\odot}$, while $M305q205_{\uparrow^{0.6} {0}}$ represents an unequal mass model, $q=2.05$, with the primary component having an aligned spin of $\chi_{1}=0.6$ relative to the orbital angular momentum, and the secondary being irrotational, with a total mass of $3.05M_{\odot}$.

\textit{Merger time ($t_{\mathrm{merger}}$) :} Merger times are determined as being the time of maximum gravitational wave amplitude, measured by a detector placed at $\approx295$km.
    
\textit{Final time :} For models resulting in a neutron star, comparisons are made at $\approx20$ms after the merger, while for models forming a black hole, comparisons are made at $\approx10$ ms after the merger.
    
\textit{Spins :} Unless otherwise specified, spins refer to the effective spin of the binary. Orientations, such as aligned or anti-aligned are defined relative to the orbital angular momentum $\mathbf{L}$. We use the following conventions:
\begin{enumerate}
        \item \textbf{irrotational} $(0 0)$, where both stars are irrotational, 
        \item \textbf{aligned} $(\uparrow \uparrow)$, or \textbf{anti-aligned} $(\downarrow \downarrow)$, where both spins are aligned (anti-aligned) with $\mathbf{L}$,
        \item \textbf{single-spin aligned} $(\uparrow 0)$ or \textbf{single-spin anti-aligned}, $(\downarrow 0)$ where the primary has aligned (anti-aligned) spin while the secondary is irrotational,
        \item \textbf{mixed} $(\downarrow \uparrow)$, where the primary has anti-aligned and the secondary has aligned spins.
\end{enumerate}

\textit{Neutrinos :} The $M0$+Leakage scheme described in Sect.~\ref{numerical_setup} is used to calculate the effect of spin on the average neutrino energies and luminosities for electron neutrinos, $\nu_{e}$, electron anti-neutrino, ${{\overline\nu}_{e}}$, and heavy-lepton neutrinos, $\nu_{x}$. Neutrino quantities are extracted at the outer boundary of the spherical grid.
    
\textit{Gravitational waves :} Gravitational waves are computed using the \texttt{WeylScal4} and \texttt{Multipole} thorns of the \texttt{Einstein Toolkit}. The \texttt{WeylScal4} thorn calculates the Newman-Penrose curvature scalar $\Psi_4$, while the \texttt{Multipole} thorn decomposes $\Psi_4$ into $s=-2$ spin-weighted spherical harmonic modes on a sphere with a radius of $\approx295$km. The strain is computed from the double time integration of $\Psi_4$ using Fixed Frequency Integration (FFI)~\citep{reisswig_notes_2011}
    \begin{equation} \label{strain}
    \begin{split}
     h &= h_+^{lm}(r,t)- i h_\times^{lm}(r,t)  \\
    & = \int_{-\infty}^t \mathrm{d}u \int_{-\infty}^u\mathrm{d}v\, \Psi_4^{lm}(r,v)
    \end{split}
    \end{equation}
    where $h_{+}$ and $h_{\times}$ represent the plus "$+$" and cross "$\times$" polarization of the gravitational waves. The quantity $\Psi_4$ is given as
    \begin{equation}\label{psi4eq}
    \Psi_4(t, r, \theta, \phi) =
    \sum_{l=2}^{l=8}\sum_{m=-l}^l \Psi_4^{lm}(t,r) {}_{-2}Y_{lm}(\theta, \phi)
    \end{equation}
The peak gravitational wave frequencies for a given $(l,m)$ mode correspond to the frequencies of the peak effective strain, $h_{\mathrm{eff}}=\sqrt{h_{+}^{2}+h_{\times}^{2}}$.
We look at energy and angular momentum loss (in $+z$  direction), frequencies and spectra of the gravitational waves, and consider modes up to $(l,m)=(8,8)$. We also discuss their detectability at a distance of $100$Mpc by the Advanced LIGO~\citep{Aasi_2015, aligo_O4high} and the Einstein Telescope (ET)~\citep{etd_sensitivity, maggiore_science_2020}. The instantaneous frequency at the moment of merger, $f_{\mathrm{merger}}$, and the post-merger peak frequencies, where $f_{1}$, $f_{2}$, and $f_{3}$ correspond to $(l,m) = (2,1), \ (2,2)$ and $(3,3)$ modes, respectively, are also presented. Among them, the $f_{2}$ frequency has been shown to correlate strongly with the radius of the maximum TOV mass, providing a direct insight into the underlying EoS~\citep{Bauswein2012, Bauswein2013, Bauswein2016}.

\textit{Ejecta :} Ejecta properties, representing unbound matter that does not fall back, are calculated from a surface located at $\sim295$km. The geodesic criteria~\citep{kastaun_properties_2015, sekiguchi2015, dietrich2015_ejecta, radice_dynamical_2016} $u_{t}<-1$, where $u_{t}$ is the time component of the four-velocity, is used. We also analyse the fast-moving component of the neutron-rich ejecta with velocity greater than $0.6$c, which is expected to yield synchrotron radiation via interaction with the interstellar medium~\citep{Hotokezeka_2018, radice_binary_2018}. The electron fraction and velocity of the matter are given as mass-weighted averages. 
    
\textit{Remnant properties :} The apparent horizon of a black hole is detected with \texttt{AHFinderDirect}~\citep{Thornburg-1996-horizon-finding, Thornburg2003:AH-finding, Brown:2008sb}. The (quasi-local) spin and mass measurement of a BH are performed using \texttt{QuasiLocalMeasures}~\citep{Dreyer:2002mx}. Following~\citet{tacik_binary_2015}, we compute the quasi-local angular momentum from $6$ spherical surfaces centred at the coordinate origin.  $5$ of these surfaces are located at radii ranging from $\approx1.48$km to $\approx16$km, where the latter approximately corresponds to the size of NS remnants. The sixth surface is placed further out, at a radius of $\approx29.5$km.
    
\textit{Disc mass :} The baryonic mass of the disc is calculated by integrating the mass within a rest-mass density threshold of $\rho < 10^{13} \, \mathrm{g cm}^{-3}$ over a radius of $\approx295$km, similarly to~\citet{radice_binary_2018} and references therein, and see also~\citet{camiletti_disk}.
If a BH is formed, we remove the region with lapse function values lower than $0.3$.
    
\section{Results for models with $M_{\mathrm{tot}}= 2.55M_{\odot}$}\label{Results}

We focus our analysis on the models with $M_{\mathrm{tot}}=2.55M_{\odot}$ and leave other models to Sect~\ref{results_other}, to improve readability.
    
This section investigates the effect of initial spin on the dynamics by analysing the inspiral times, listed in Table~\ref{initial-data}, and identifying possible mechanisms responsible for observed differences and variations.

\begin{figure*} 
        \centering
    
        \includegraphics[width=2.0\columnwidth]{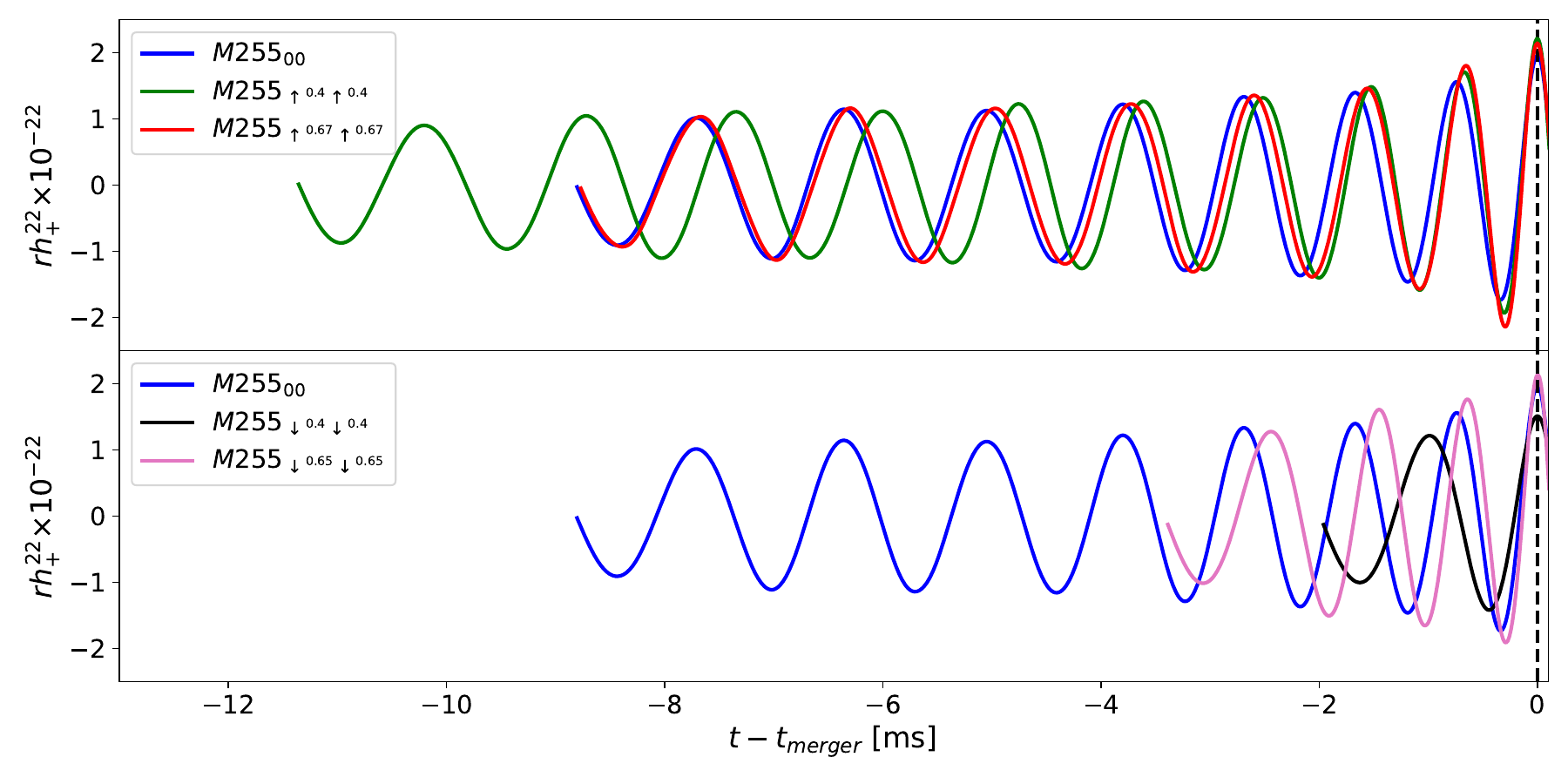}
        \caption{Gravitational wave strains for aligned (top panel) and anti-aligned (bottom panel) spin models, shown up to $0.1$ ms after the merger from high resolution simulations. The strains are aligned at the merger time (vertical dashed line), with the change in trend with spin distinctly visible in both panels: note differences in merger times.}
        \label{fig:gw-uzak}
    \end{figure*}
    
    \textbf{Impact of spin on the inspiral phase: } As discussed in Sect.~\ref{introduction}, previous studies have shown that spin-orbit interaction is repulsive, resulting in delay of merger, for aligned spins and attractive for anti-aligned spins, producing a speed-up of merger. Our simulations, however, reveal a change of this qualitative trend. It arises probably from the interplay between spin-orbit and spin-spin interactions, for spins in the range of $\chi=0.4-0.67$ in both aligned and anti-aligned models. In this regime, increasing the spin decreases inspiral time for aligned spins; and vice-versa for anti-aligned spins, inpsiral time increases. The single-spin models have only $|\chi_1|=0.4$ and hence are outwith the parameter range. These follow the general trend of previous studies. 
    
    The mixed spin models further illustrate the importance of individual spins for the inspiral time. Despite having $\chi= 0$, $M255_{\downarrow^{0.4} \uparrow^{0.4}}$ and 
$M255_{\downarrow^{0.65} \uparrow^{0.65}}$ merge $\sim 4.9$ and $\sim 2.4$ ms earlier than the irrotational model. Increasing the individual spin from $|\chi_{1,2}| = 0.4$ to $0.65$ therefore shortens the inspiral, consistent with the change in trend observed in aligned and anti-aligned models. 

Since the equatorial bulge increases with spin, two aligned spin models, $M255_{\uparrow^{0.4} \uparrow^{0.4}}$ $M255_{\uparrow^{0.67} \uparrow^{0.67}}$, are simulated with a larger initial separation of $\sim60$km. These models are not included in any analysis and do not appear in any figure or table. They serve solely as a robustness check, confirming that the observed change in trend is independent of the initial separation. They exhibit a similar trend regarding the time spent in the inspiral phase as compared to other models, allowing us to rule out any significant impact of the initial separation on the observed change in trend with spin.

    A similar change in trend is identified for anti-aligned spins between $M255_{\downarrow^{0.4} \downarrow^{0.4}}$ and $M255_{\downarrow^{0.65} \downarrow^{0.65}}$. Although our simulations for larger separation are focused on aligned spins, we hypothesize that this behaviour in anti-aligned spins follows the same underlying mechanism, the interplay between spin-orbit and spin-spin interactions. To support this interpretation, Fig.~\ref{fig:gw-uzak} presents the gravitational wave strain until $0.1$ ms after the merger for both aligned and anti-aligned spins alongside the irrotational model. The figure highlights the change in trend across different spin configurations. 
    
    For moderate spins of $\chi=0.4$, spin-orbit interaction largely dominates the dynamics, leading to a longer inspiral phase. However, as the spin increases beyond $\chi=0.4$ spin-spin effects become increasingly important, counteracting the spin-orbit interactions~\citep{Dietrich_2017_rotation}, causing earlier (later) mergers despite higher spin for aligned (anti-aligned) spins. While this behaviour is observed consistently at both low and high resolutions (see Table~\ref{initial-data}), we note that its robustness at even higher resolutions may require further confirmation.

    \textbf{Evolution of Thermodynamic Properties: } We now show the impact of the spin effects and differences on the maximum rest-mass density and temperature of the matter. 
    
    For models with spins up to $|\chi|=0.4$, aligned (anti-aligned) spins result in a longer (shorter) inspiral phase and less (more) violent mergers, as evidenced by reaching lower (higher) maximum temperatures and rest-mass densities as compared to the irrotational model. For higher spins, such as $M255_{\downarrow^{0.65} \downarrow^{0.65}}$ and $M255_{\uparrow^{0.67} \uparrow^{0.67}}$, despite the change in trend observed in orbital dynamics, the overall trend remains consistent: increasing aligned (anti-aligned) spin continues to produce less (more) violent mergers. The values of maximum temperatures and rest-mass densities change with spin peaking at $\sim145$ MeV and $\sim5.6\rho_{\mathrm{sat}}$ for the $M255_{\downarrow^{0.65} \downarrow^{0.65}}$ model, where $\rho_{\mathrm{sat}}$ is the nuclear saturation density of $2.7\times10^{14}\, \mathrm{g cm}^{-3}$~\citep{lattimer_nuclear_2012}. This suggests that for spins above $|\chi|=0.4$, where both spin-orbit and spin-spin interactions significantly influence the dynamics, merger intensity cannot be solely determined by the timing of the merger.

    \begin{figure} 
        \centering
    
        \includegraphics[width=1.0\columnwidth]{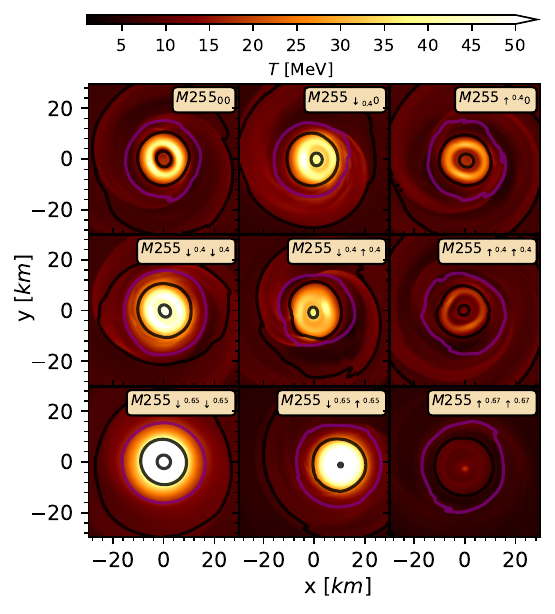}
        \caption{
    The temperature distribution of the remnant and the inner disc region is shown for $M_{\mathrm{tot}}=2.55M_{\odot}$ at $20$ ms after the merger in the $x$-$y$ plane. The purple contour marks a rest-mass density of $\rho=10^{13} \, \mathrm{g cm}^{-3}$, black contours denote $\rho=10^{12}$, $10^{14}$ and $10^{15} \, \mathrm{g cm}^{-3}$. Panels represent the different spin configurations, illustrating the impact of spin on the temperature structure of the remnant.
    }
        \label{fig:temp-m255-yakin}
    \end{figure}

    \begin{figure}  
        \centering
        
        \includegraphics[width=1.0\columnwidth]{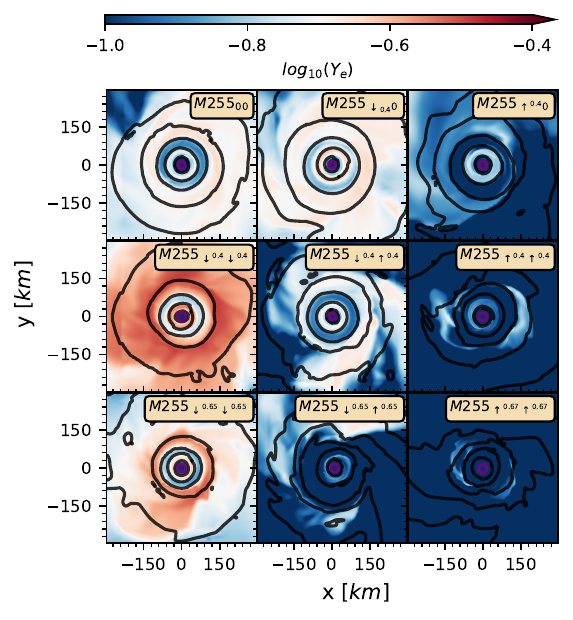}
         \caption{The distribution of electron fraction $Y_{e}$ in the remnant and disc is shown at $20$ ms after the merger. Both $Y_{e}$ and the rest-mass density contours are presented on logarithmic scales, with black contours marking densities of $\rho= 10^{6}, \, 10^{7}, \, 10^{8}, \, 10^{9}, \, 10^{10}, \, 10^{11}, \, 10^{12} \, \mathrm{g cm}^{-3}$, and the purple contour representing $\rho = 10^{13} \, \mathrm{g cm}^{-3}$. This figure highlights the impact of spin on the composition.}
        \label{fig:ye-hmns-uzak}
    \end{figure}

Snapshots of the temperature of the remnant and of the composition of the disc, along with the rest-mass density contours, at $20$ms after the merger, are shown in Fig.\ref{fig:temp-m255-yakin} and in Fig.\ref{fig:ye-hmns-uzak}, respectively. These figures illustrate that in aligned (and single-spin aligned) models the additional angular momentum increases the rotational support of the remnant, making it less compact and reducing shock heating, which results in a cooler core surrounded by a hot envelope with the maximum densities reached more gradually, following the same pattern as the irrotational model.
The redistribution of the additional angular momentum leads to the formation of spiral arms the strength of which increases with spin. At $20$ ms after the merger, only part of the remnant of these arms are visible for the aligned spin model $M255_{\uparrow^{0.67} \uparrow^{0.67}}$. 

In contrast, for anti-aligned (including single-spin anti-aligned) models, the reduced angular momentum weakens the rotational support of the remnant, leading to a more compact and hotter remnant with densities peaking almost immediately after the merger. In mixed spin models, enhanced spin-spin interaction results in a more violent
merger. It leads to strong shock heating and a uniformly hot core for $M255_{\downarrow^{0.65} \uparrow^{0.65}}$ and a hotter core than irrotational model despite weaker heating for $M255_{\downarrow^{0.4} \uparrow^{0.4}}$. These features are illustrated in Figure~\ref{fig:temp-m255-yakin}.
    
In the single-spin models, $M255_{\downarrow_{0.4}0}$, and $M255_{\uparrow^{0.4} 0}$, the spinning neutron stars become tidally disrupted, due to larger equatorial bulge, but in the aligned spin model the tidal tail is more pronounced due to the additional angular momentum.
Looking at the impact of spin on the compositional change of the disc, we see in Fig.~\ref{fig:ye-hmns-uzak} that the mixed spin model, $M255_{\downarrow^{0.65} \uparrow^{0.65}}$, still retains a tidal stream which is mainly composed of neutrons. In general, anti-aligned (aligned) spins result in less (more) neutron-rich discs compared to the irrotational models. This trend can be attributed to the impact of weak interactions, which are discussed in the following section.
    
\textbf{Neutrinos: } Neutrinos play a crucial role in determining the composition of matter through weak interactions and influence the stability and thermodynamic properties of the remnant neutron star by carrying energy away. They also drive the ejecta to higher altitudes (larger $|z|$ values) and affect its composition~\citep{radice_binary_2018}. These quantities, along with the mass and velocity of the ejecta, determine r-process nucleosynthesis and properties of the ensuing kilonovae. See~\citet{radice_binary_2018, Espino_neutrino2024} for the impact of neutrinos on the ejecta and~\citet{Foucart_review2023} for a recent review of the impact of neutrinos in BNS mergers. 
    
\begin{figure*} 
    \centering
    
    \includegraphics[width=2.1\columnwidth]{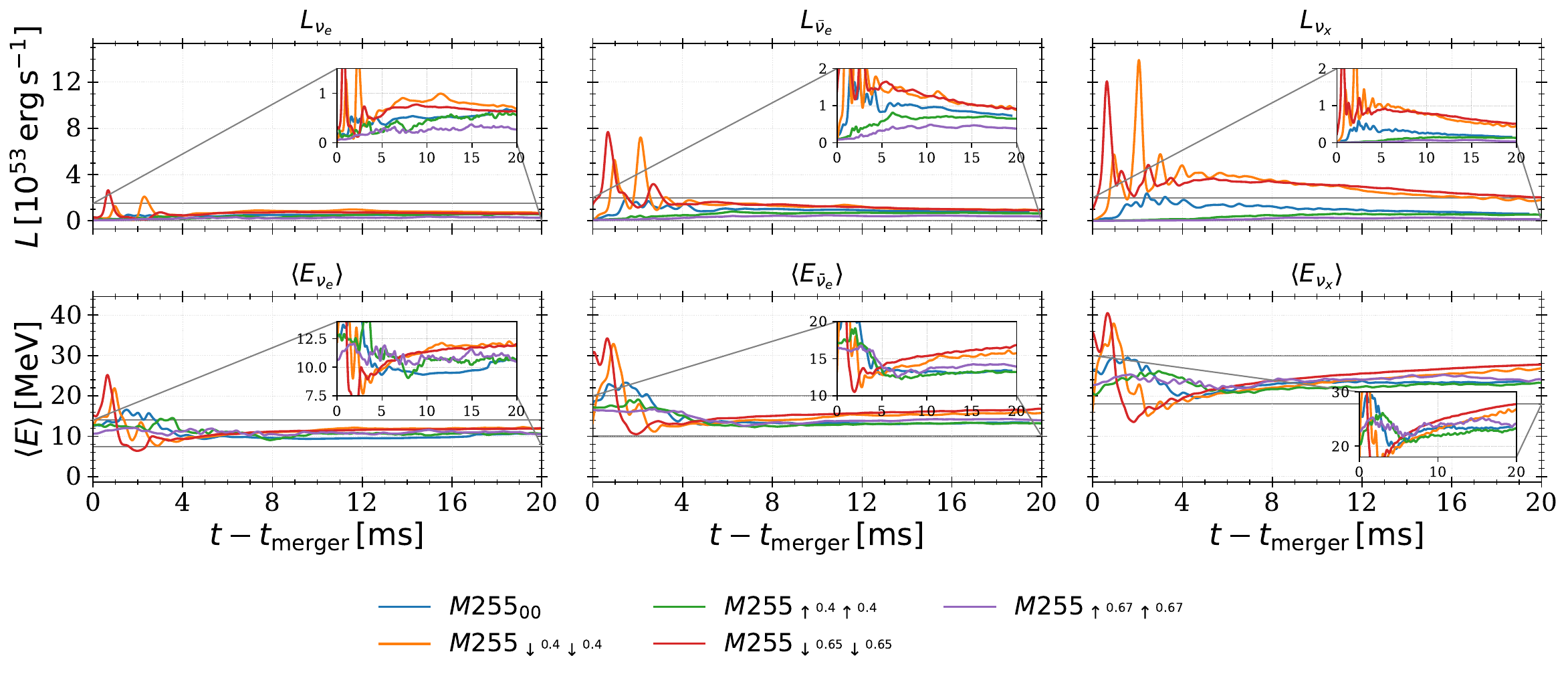}
    \caption{
Neutrino luminosities and mean energies for electron neutrinos ($\nu_e$), electron anti-neutrinos ($\bar{\nu}_e$), and heavy‐lepton neutrinos ($\nu_x$) for $M_{\mathrm{tot}}=2.55\,M_{\odot}$ models with different spins. Anti-aligned spins yield higher mean energies and enhanced luminosities than the aligned spins due to higher neutrino number flux from more compact and hotter remnants. Main panels share a common y-axis to illustrate overall trends, whereas inset panels use their own scales to enhance visualisation of temporal variations.
    }
        \label{fig:neutrino}
\end{figure*}

Using the M0+Leakage scheme, details of which can be reviewed in Sect.~\ref{numerical_setup}, we now discuss the impact of initial spin on average neutrino energies and luminosities for the electron neutrino ($\nu_{e}$), electron anti-neutrino (${{\overline\nu}_{e}}$) and the heavy-lepton neutrinos ($\nu_{x}$), as presented in Fig.~\ref{fig:neutrino}. 
This figure enables a direct comparison of mean energies and luminosities for all flavours, covering aligned and anti-aligned spin configurations as well as the irrotational model. In all models, the flavour hierarchy remains consistent: $\langle E_{\nu_x} \rangle > \langle E_{\bar{\nu}_e} \rangle > \langle E_{\nu_e} \rangle$ and this does not change across any spin configurations considered here. This ordering is consistent with~\citet{foucart_neutrino2016, Radice_M1_2022, radice_neutrino2023}.

In contrast, the luminosity hierarchy varies with spin. Anti-aligned models produce compact, hot remnants that enhance heavy-lepton neutrino emission, resulting in a hierarchy of $L_{\nu_x} > L_{\bar{\nu}_e} > L_{\nu_e}$, similar to the irrotational and single-spin anti-aligned model. Despite being less compact than these models, and experiencing stronger shock heating than the irrotational model, the mixed spin model $M255_{\downarrow^{0.65} \uparrow^{0.65}}$ also exhibits the same hierarchy.
Here $L_{\nu_x}$ denotes the total luminosity of the heavy-lepton neutrinos, the sum over $\nu_\mu$, $\overline{\nu}_{\mu}$, $\nu_\tau$ and $\overline{\nu}_{\tau}$.

Aligned spin models by contrast, yield more extended and dilute remnants with lower temperature, which suppress the overall neutrino number flux. In these models, the dominant emission shifts to electron anti-neutrinos, leading to a luminosity hierarchy of $L_{\bar{\nu}_e} > L_{\nu_e} > L_{\nu_x}$. On the other hand, single-spin aligned and mixed spin model $M255_{\downarrow^{0.4} \uparrow^{0.4}}$ show enhanced $\nu_x$ relative to $\nu_e$ leading to the hierarchy of $L_{\bar{\nu}_e} > L_{\nu_x} > L_{\nu_e}$. 

These results highlight that spin orientation strongly affects the morphology and thermal properties of the remnant, which in turn shapes the neutrino emission properties.

\begin{table*}
\centering
\caption{Energy and angular momentum radiated by gravitational waves, along with merger and peak frequencies, of modes contributing more than $2\times10^{-3}$ to the total energy release. $E^{\mathrm{gw}}_{\mathrm{ins}}$ ($J^{\mathrm{gw}}_{\mathrm{ins}}$), $E^{\mathrm{gw}}_{\mathrm{tot}}$ ($J^{\mathrm{gw}}_{\mathrm{tot}}$) and $E^{\mathrm{gw}}_{\mathrm{tot}}/M_{\mathrm{tot}}^{\mathrm{loc}}$ ($J^{\mathrm{gw}}_{\mathrm{tot}}/J_{0}$) show energy (angular momentum) loss during inspiral, the total energy (angular momentum) loss and the ratio of the total energy (angular momentum) loss to the total initial mass energy (angular momentum), respectively. Additionally, $f_\mathrm{merger}$ denotes the instantaneous frequency at the merger time, while $f_1$, $f_2$, and $f_3$ represent the peak frequencies corresponding to the $(l,m) = (2,1)$, $(2,2)$, and $(3,3)$ modes, respectively. For remnants undergoing prompt collapse to a black hole, only $f_\mathrm{merger}$ is provided. When parentheses are present, the values in parentheses correspond to low-resolution simulations, while non-parenthesized values represent high-resolution simulations.}

\small
    
\begin{tabular}{p{1.4cm}*{10}{>{\centering\arraybackslash}p{1.15cm}}}
    
    \toprule
    Model & $E^{\mathrm{gw}}_{\mathrm{ins}}$ ~~~ (\%) & $E^{\mathrm{gw}}_{\mathrm{tot}}$ ($\times10^{52}$ ergs)  & $E^{\mathrm{gw}}_{\mathrm{tot}}/M_{\mathrm{tot}}^{\mathrm{loc}}$ $(\times10^{-2})$ & $J^{\mathrm{gw}}_{\mathrm{ins}}$ ~~~ (\%) &$J^{\mathrm{gw}}_{\mathrm{tot}}$ ($M^{2}_{\odot})$ & $J^{\mathrm{gw}}_{\mathrm{tot}}/J_{0}$ $(\times10^{-2})$ & $f_{\mathrm{merger}}$ (kHz) & $f_{1}$ ~~~~ (kHz) & $f_{2}$ ~~~~ (kHz) & $f_{3}$ ~~~~ (kHz) \\ 
    \midrule
    $M255_{00}$ &$20(22)$&$9.18(9.64)$ &$2.04(2.14)$& $42(43)$&$1.59(1.64)$ &$25.17(25.83)$ & $1.70 (2.08)$ & $1.46(1.54)$ & $2.89(2.93)$& $4.35(4.34)$ \\
    $M255_{\downarrow_{0.4}0}$  & $8(12)$ &$8.45(5.45)$ &$1.87(1.21)$& $19(27)$&$1.25(0.86)$ &$21.50(14.88)$ & $1.34(1.41)$ & $1.46(1.48)$ & $2.91(2.99)$& $4.26(4.32)$\\
    $M255_{\uparrow^{0.4} 0}$  &$23(24)$ &$8.54(9.49)$ &$1.89(2.10)$& $46(46)$&$1.58(1.68)$ &$22.61(23.97)$ & $1.62(2.03)$ & $1.51(1.53)$ & $3.03(2.97)$& $4.38(4.44)$\\
    $M255_{\downarrow^{0.4} \downarrow^{0.4}}$  &$14(9)$ &$3.16(5.03)$ &$0.70(1.11)$& $29(19)$&$0.54(0.80)$ &$10.39(15.32)$ & $1.32(1.34)$  & $1.46(1.44)$ & $2.80(2.69)$& $4.14(4.05)$\\
    $M255_{\downarrow^{0.4} \uparrow^{0.4}}$  &$27(17)$ &$4.68(7.68)$ &$1.03(1.69)$& $46(33)$&$0.83(1.16)$ &$12.85(18.03)$ & $2.09(2.23)$ &  $1.48(1.54)$ & $2.86(3.04)$& $4.24(4.51)$ \\
    $M255_{\uparrow^{0.4} \uparrow^{0.4}}$   &$32(39)$ &$7.74(7.60)$ &$1.71(1.68)$& $55(61)$&$1.55(1.49)$ &$20.39(19.67)$ & $1.90(2.38)$ &  $1.47(1.49)$& $3.01(2.97)$& $4.51(4.54)$\\
    $M255_{\downarrow^{0.65} \downarrow^{0.65}}$  &$30(38)$ &$5.80(4.83)$ &$1.29(1.07)$& $45(56)$&$1.09(0.89)$ &$24.32(19.95)$ & $2.01(2.24)$ &  $2.54(1.21)$ & $2.50(2.56)$&$2.54(3.73)$\\
    $M255_{\downarrow^{0.65} \uparrow^{0.65}}$   &$67(72)$ &$2.31(2.32)$ &$0.51(0.52)$& $77(82)$&$0.70(0.68)$ &$10.95(10.67)$ & $1.56(1.75)$ & $2.58(2.53)$ & $2.62(2.53)$& $2.55(2.56)$\\
    $M255_{\uparrow^{0.67} \uparrow^{0.67}}$  &$69(51)$ &$3.35(3.33)$ &$0.75(0.75)$& $83(68)$&$0.92(0.90)$ &$11.01(10.72)$ & $1.93(1.55)$ & $1.49(1.48)$ & $3.00(2.96)$&$4.22(4.33)$ \\
    $M305_{{0}{0}}$  &$57$ &$5.50$ &$1.02$& $75$&$1.34$ &$15.46$ & $1.94$ & - & -  &-\\
    $M305_{\downarrow^{0.4} \downarrow^{0.4}}$  &$26$ &$3.89$ &$0.72$& $50$&$0.64$ &$9.07$& $1.66$ & - & -\\
    $M305_{\uparrow^{0.4} \uparrow^{0.4}}$  &$39$ &$13.87$ &$2.57$& $63$&$2.32$ &$22.48$ & $2.70$ & $3.88$ &3.60&5.08\\
    $M305_{\uparrow^{0.67} \uparrow^{0.67}}$  &$62$ &$5.29$ &$0.98$& $76$&$1.38$ &$11.99$ & $1.87$ & $1.68$ &3.14&4.61\\
    $M305q205_{{0}{0}}$  &$69$ &$2.27$ &$0.42$& $77$&$0.74$ &$9.74$ &$1.24$ & - & -&- \\
    $M305q205_{\uparrow^{0.6} {0}}$  &$60$ &$2.80$ &$0.52$& $75$&$0.85$ &$8.53$ & $1.39$ & $3.68$ & 3.58 & 5.23\\
    $M410_{{0} {0}}$  &$64$ &$19.47$ &$2.69$& $79$&$3.31$ &$22.93$ &$3.17$ &  - & -&-\\
    $M410_{\downarrow^{0.65} \downarrow^{0.65}}$  &$59$ &$11.01$ &$1.52$& $73$&$1.82$ &$18.78$ & $2.96$  & - & -&-\\
    $M410_{\downarrow^{0.65} \uparrow^{0.65}}$   &$63$ &$13.92$ &$1.93$& $76$&$2.54$ &$17.62$ &$2.77$  & - & -&-\\
    $M410_{\uparrow^{0.67} \uparrow^{0.67}}$  &$75$ &$16.04$ &$2.22$& $85$&$3.43$ &$17.83$ & $2.66$  & - & -&-\\
    \hline
    
\label{gw_energy}
\end{tabular}
\end{table*}
    
\textbf{Gravitational waves: } The impact of spin on gravitational wave emission is analysed by examining the total extracted energy and angular momentum loss, their relative contributions during the inspiral phase, and the ratio of total energy and angular momentum release to the initial values for mass-energy and angular momentum. In Table~\ref{gw_energy} we summarise these quantities, along with the merger and peak frequencies of modes contributing more than $2\times10^{-3}$ to the total energy release. We stress that gravitational wave quantities during the inspiral phase are meaningful only within the context of an initial separation of $40$ km, and do not represent absolute physical differences across models.
    
Models with aligned spins radiate more energy and angular momentum than anti-aligned spin models with the same spin magnitude, while the irrotational model exhibits the highest overall energy and angular momentum release. The gravitational wave strains, shown in Fig.~\ref{fig:gw-uzak}, clearly demonstrate the impact of spin on the dynamics. During the post-merger phase (not shown in Fig.~\ref{fig:gw-uzak}), spin magnitude and orientation significantly influence the remnant oscillation, mimicking the effects of softer or stiffer EoS or unequal mass binaries. Gravitational wave radiation efficiency decreases with increasing anti-aligned spin due to the formation of a more compact remnant. Conversely, aligned spins lead to less compact remnants with extended spiral arms, enhancing the radiation efficiency. Single-spin models follow the aligned (anti-aligned) models, with differences in energy and angular momentum radiation and frequency shifts in $f_1$ and $f_2$ remaining within the estimated numerical uncertainties, and thus not indicative of new behaviour, while $M255_{\downarrow^{0.65} \uparrow^{0.65}}$ model shows the lowest energy emission through gravitational waves.
    
\begin{figure}
    
         \includegraphics[width=1.0\columnwidth]{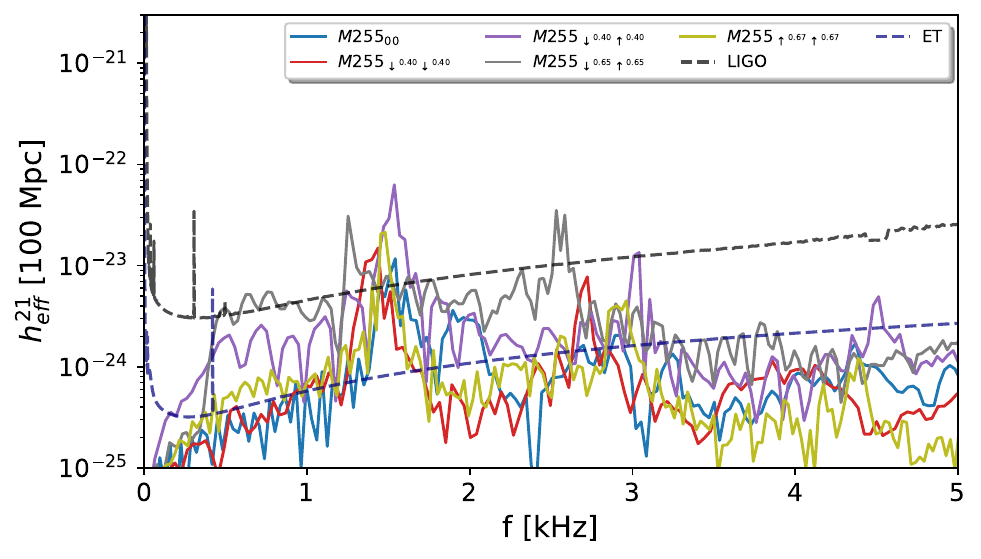}
        \caption{
        Gravitational wave spectra for mode $(l,m)=(2,1)$ for models with $M_\mathrm{{tot}}=2.55M_{\odot}$ showing their detectability by the Advanced LIGO and the Einstein Telescope (ET). The dotted lines indicate the sensitivity curves of the detectors \citep{etd_sensitivity, aligo_O4high}}.
        \label{fig:gw_detectabilityl2m1}
    \end{figure}

\begin{figure}
    
      \includegraphics[width=1.0\columnwidth]{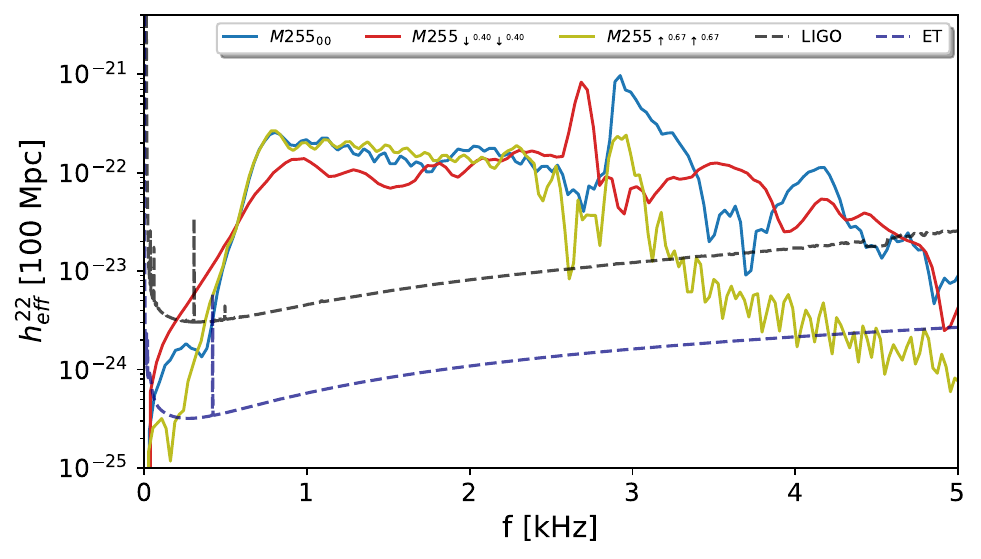}
      \caption{
      A similar presentation as in Fig.\ref{fig:gw_detectabilityl2m1} but for the fundamental mode, $(l,m)=(2,2)$, focusing on highly aligned, anti-aligned spin and irrotational models.
      }
    \label{fig:gw_detectabilityl2m2}
    \end{figure}

In Figs.~\ref{fig:gw_detectabilityl2m1} and~\ref{fig:gw_detectabilityl2m2} we show the gravitational wave spectra for the $(2,1)$ and $(2,2)$ modes, along with the sensitivity curves of Advanced LIGO and the Einstein Telescope (ET). We observe that the peak frequency of the fundamental mode $f_2$, shifts to higher frequencies for aligned spins, consistent with~\citet{Dietrich_2017_rotation}, but in contrast to~\citet{bernuzzi_mergers_2014, East_2019, rosswog_2024}. For anti-aligned spins, the frequency shift is more pronounced, with the largest difference compared to the irrotational model being $\sim0.39$ kHz for $M255_{\downarrow^{0.65} \downarrow^{0.65}}$. These shifts are larger than the estimated numerical uncertainties, which we obtain by comparing high and low resolution values with $\Delta f = |f_{\mathrm{HR}} - f_{\mathrm{LR}}|$, at most $ \Delta f_2 \lesssim 0.18$ kHz. Shifting to the lower frequencies for the anti-aligned spins contrasts with~\citet{East_2019}, who considered $\chi=-0.13$, likely due to the higher spins, $\chi=-0.4$ and $\chi=-0.65$, analysed in this study. These shifts in the $f_2$ demonstrate that spins, particularly high spins (both aligned and anti-aligned), can introduce degeneracies in the gravitational wave spectra, complicating the inference of the EoS. A shift to lower (higher) frequencies mimics the effects of a stiffer (softer) EoS, compared to the irrotational model. Unlike $f_2$ frequencies, the $(2,1)$ mode, associated with the one-arm spiral instability shows shifts within numerical uncertainties, found to be at most $\Delta f_1 \lesssim 0.08$ kHz, except for $M255_{\downarrow^{0.65} \downarrow^{0.65}}$, where the difference reaches $\sim 1.23$ kHz.

\textbf{Ejecta: } In Table~\ref{ejecta_properties_table} we
summarise the properties of the ejected matter, including total mass, fast-moving ejecta mass, velocity and electron fraction. Across all spin configurations, the total ejecta mass increases compared to the irrotational model, with aligned spins producing a higher total ejecta mass than the anti-aligned spins. Single-spin models yield ejecta masses broadly consistent with aligned (anti-aligned) models, while the mixed spin model $M255_{\downarrow^{0.65} \uparrow^{0.65}}$ produces an ejecta mass comparable to that of the $M255_{\uparrow^{0.67} \uparrow^{0.67}}$ model, both give $\sim0.06M_{\odot}$.

The composition shows a clear dependence of spin orientation: anti-aligned spins generally result in less neutron-rich ejecta compared to the aligned spins.
    
The mass of the fast-moving ejecta, as defined in Sect.~\ref{Analysis}, lies between $(10^{-8}-10^{-4})M_{\odot}$, and depends on the EoS, as demonstrated in~\citet{radice_binary_2018}. Specifically, binaries with SFHo EoS were shown to exhibit fast-moving ejecta not only during the merger, but also after the first bounce of the remnant, a behaviour unique among the EoSs considered. In our study, we observe fast-moving ejecta in the anti-aligned spin models (including single-spin anti-aligned), driven by the strong shock heating, which is notably absent in the aligned spin models, and is strongly suppressed in the mixed spin models. The calculated fast-moving ejecta masses align with those reported in~\citet{radice_binary_2018}.
    
\begin{table*}
    \caption{The ejecta properties and disc masses at final times; see Sect.\ref{Analysis} for details. The ejecta properties are extracted at a distance of
    $r=295$ km. $M_\mathrm{ej}$ and $M_{\mathrm{ej}}\geq0.6$c denote the mass of total ejecta, and of 
    fast-moving ejecta, respectively. $M_{\mathrm{disc}}$ represents the disc mass; $\langle v \rangle$ and $\langle Y_{e} \rangle$ correspond to the mass-weighted average velocity and electron fraction, respectively. When parentheses are present, the values in parentheses correspond to low-resolution simulations, while non-parenthesized values represent high-resolution simulations.
    }
    
    \begin{tabular}{@{}lccccccccccc@{}}
    \hline
    \small
    Model & 
    \multicolumn{1}{c}{\makecell{$M_{\mathrm{ej}}$ \\ $(10^{-3} M_{\odot})$}} & 
    \multicolumn{1}{c}{\makecell{$M_{\mathrm{ej}} \geq 0.6c$ \\ $(10^{-5} M_{\odot})$}} & 
    \multicolumn{1}{c}{\makecell{$M_{\mathrm{disc}}$ \\ $(10^{-2} M_{\odot})$}} & 
    \multicolumn{1}{c}{\makecell{$\langle Y_{e} \rangle$}} & 
    \multicolumn{1}{c}{\makecell{$\langle v \rangle$ \\ $(c)$}} \\
    \midrule

    $M255_{00}$&1.83(1.24) & 0.05 (0.00) & 24 (19) & 0.32 (0.33) & 0.21 (0.21)  \\
    $M255_{\downarrow_{0.4}0}$&3.73 (6.86) & 0.69 (2.72) & 14 (21) & 0.31 (0.31) & 0.22 (0.22) \\
    $M255_{\uparrow^{0.4} 0}$&2.25 (2.93) & 0.00 (0.00) & 16 (29) & 0.31 (0.41) & 0.22 (0.22)  \\
    $M255_{\downarrow^{0.4} \downarrow^{0.4}}$&6.64 (11.64) & 8.40 (17.36) & 16 (13) & 0.24 (0.25) & 0.23 (0.24)  \\
    $M255_{\downarrow^{0.4} \uparrow^{0.4}}$&2.33 (3.65) & 0.06 (0.58) & 22 (16) & 0.32 (0.30) & 0.22 (0.22)     \\
    $M255_{\uparrow^{0.4} \uparrow^{0.4}}$&11.27 (14.59) & 0.00 (0.00) & 38 (37) & 0.27 (0.56) & 0.24 (0.23)   \\
    $M255_{\downarrow^{0.65} \downarrow^{0.65}}$&17.01 (11.78) & 28.42 (42.64) & 22 (18) & 0.21 (0.19) & 0.24 (0.24)   \\
    $M255_{\downarrow^{0.65} \uparrow^{0.65}}$&57.33 (55.45) & 0.00 (0.11) & 20 (23) & 0.30 (0.32) & 0.21 (0.21) \\ 
    $M255_{\uparrow^{0.67} \uparrow^{0.67}}$&58.73 (54.85) & 0.00 (0.00) & 28 (28) & 0.29 (0.37) & 0.22 (0.21)  \\
    $M305_{{0}{0}}$&0.38 &  9.30 & 0.002 &0.36 &0.43  \\
    $M305_{\downarrow^{0.4} \downarrow^{0.4}}$&0.65 &  2.47 & 0.2 &0.35 &0.26  \\
    $M305_{\uparrow^{0.4} \uparrow^{0.4}}$ &8.14 &  0.00 & 16.57 &0.35&0.26     \\
    $M305_{\uparrow^{0.67} \uparrow^{0.67}}$ &57.16 &  0.00 & 25.93 &0.35&0.23     \\
    $M305q205_{{0}{0}}$&10.68 &  24.23 & 11.22 &0.40 &0.31   \\
    $M305q205_{\uparrow^{0.6} {0}}$&17.31 &  0.32 & 27.09 &0.36 &0.22   \\
    $M410_{{0} {0}}$&0.02 &  0.00 & 0.001&0.38 &0.27  \\
    $M410_{\downarrow^{0.65} \uparrow^{0.65}}$&17.47&  4.10 & 6.25 &0.37 &0.35    \\
    $M410_{\downarrow^{0.65} \downarrow^{0.65}}$&0.00  &  0.00 & 0.00 &0.00&0.00    \\
    $M410_{\uparrow^{0.67} \uparrow^{0.67}}$&46.94 &  0.04 & 23.16 &0.47 &0.28   \\
    \hline
    
    \label{ejecta_properties_table}
    
    \end{tabular}
    \end{table*}
    
\begin{figure}
    \centering
     \includegraphics[width=1.0\columnwidth]{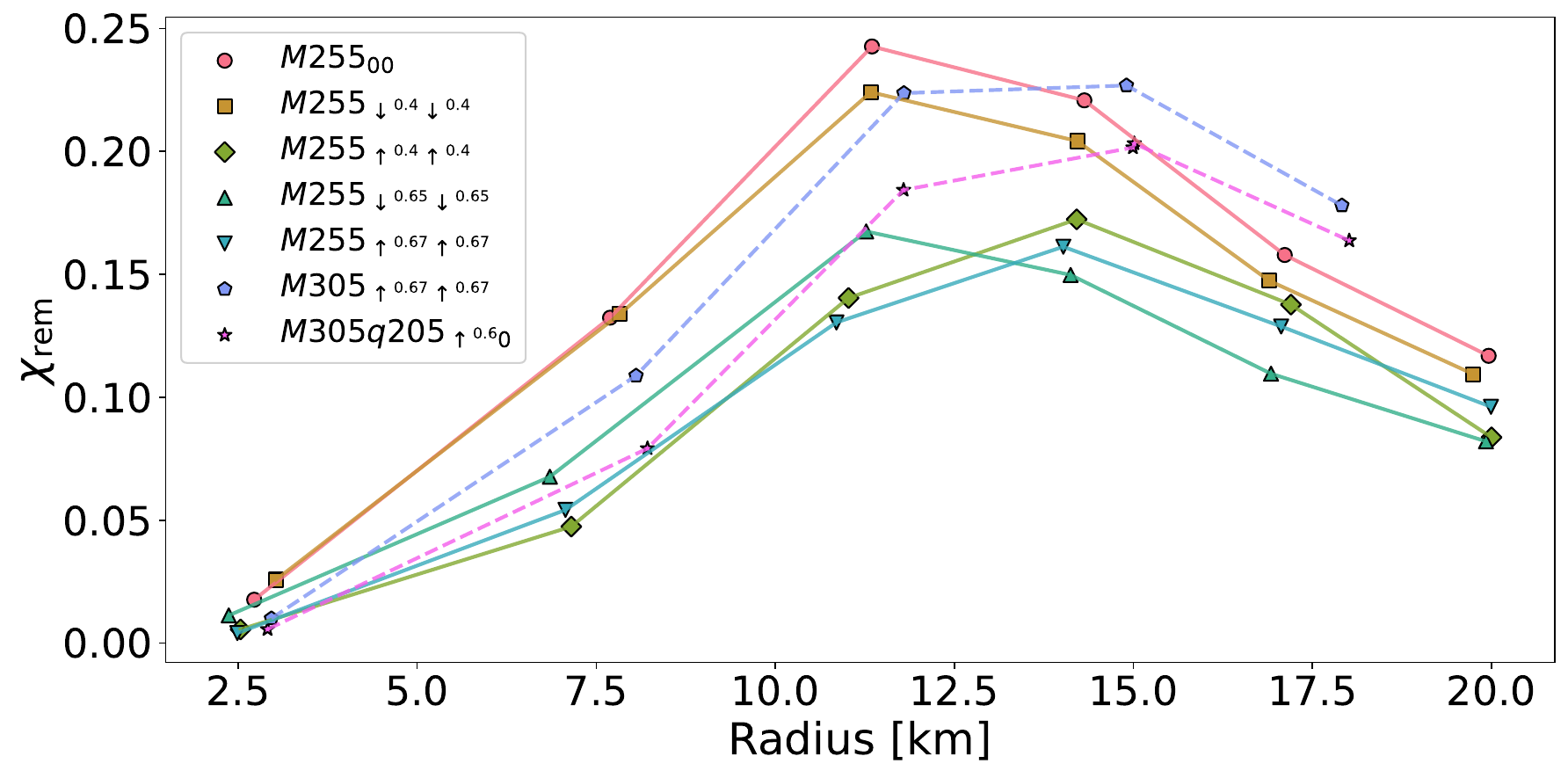}
    \caption{
    The quasi-local dimensionless spin of the post-merger remnant as a function of radius for models that resulting in a long-lived neutron star. Dashed lines represent models with $M_{\mathrm{tot}}=3.05M_{\odot}$, showing the impact of total mass and mass asymmetry on the spin profile.
     }
    \label{fig:bns-postmerger-spin}
\end{figure}

\textbf{Remnants: } All remnants with $M_{\mathrm{tot}}=2.55M_{\odot}$ result in long-lived neutron stars. The rotation pattern of the BNS remnants differs from single, differentially rotating neutron star, where the core rotates faster than the envelope. In contrast, BNS merger remnants exhibit an inverse rotation pattern, where the core rotates more slowly than the envelope~\citep{shibata2006, Kastaun_stable, East_2019}. This behaviour is illustrated in Fig.~\ref{fig:bns-postmerger-spin} which shows the variation of the dimensionless spin parameter of the long-lived neutron star across different radii. 

Among all models, the highest spin is observed in the irrotational model, with $\chi_{\mathrm{rem}}=0.24$. While this may initially appear counter-intuitive, it is a direct consequence of angular momentum dynamics. As shown in Table~\ref{initial-data}, anti-aligned spin models begin with significantly lower total angular momentum ($J_0$) because the spin vectors are oriented in the opposite direction to the orbital angular momentum and therefore reduce the total angular momentum through spin-orbit interactions. These models also produce slightly more massive remnants, contributing to a further reduction in spin. In contrast, aligned spins start with higher $J_0$. However, during the merger, the formation of extended spiral arms leads to a broader redistribution of angular momentum. Consequently, a smaller fraction of the initial angular momentum is retained in the remnant where the spin is measured, resulting in a lower spin despite the higher $J_0$.

Aligned spins also produce remnants with lower rest-mass densities and more extended structures, whereas anti-aligned models yield compact, high-density remnants. Prior to merger, neutron stars in anti-aligned spins undergo significant elongation, likely influenced by a combination of tidal, spin-orbit and spin-spin interactions. In contrast, aligned spin models have increased equatorial bulge and are therefore more prone to tidal disruption. This is particularly evident in the $M255_{\downarrow^{0.65} \uparrow^{0.65}}$ model, which exhibits the most pronounced tidal tail. The results demonstrate that spin magnitude and orientation play a key role in determining the likelihood of tidal disruption, hence the post-merger remnant.

\begin{figure}  
    \centering
    
    \includegraphics[width=1.0\columnwidth]{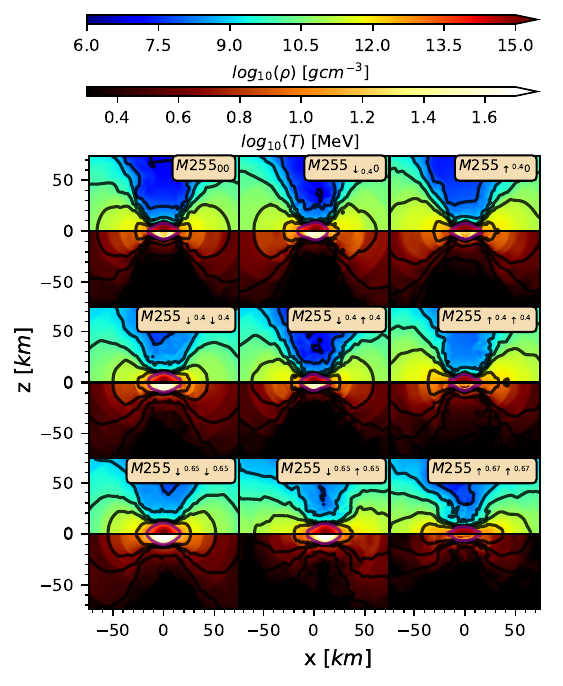}    
    \caption{Temperature and rest-mass density distribution in the $x$-$z$ plane, showing the disc structure at $20$ ms after the merger for models with $M_{\mathrm{tot}}= 2.55M_{\odot}$. The rest-mass density contours are the same as in Fig.~\ref{fig:ye-hmns-uzak}.
     }
    \label{fig:hmns-disk-rhor100}
\end{figure}

\textbf{Disc mass: } The rest-mass density and temperature of the disc in the $x$-$z$ plane, shown in Fig.~\ref{fig:hmns-disk-rhor100}, illustrate the impact of spin on these properties. Aligned spin models exhibit a more neutron-rich disc compared to both irrotational and anti-aligned spin models. This enhanced neutron abundance is generally associated with higher opacities and dimmer kilonovae~\citet{kasen_opacities_2013}. However, since we do not perform radiative transfer or light curve analysis in this study, we refrain from drawing quantitative conclusions about the electromagnetic counterpart. These aspects will be explored in future work.
The disc masses are summarized in Table~\ref{ejecta_properties_table}, with their dependence on the effective spin shown in Fig.~\ref{fig:spindisc_fit}. The figure suggests a correlation between effective spin and disc mass, with linear regression fits derived using the least squares method. While deviations from the fits are evident, possibly due to the low resolution, the overall trend highlights the importance of the impact of the effective spin on the disc mass. For $M_{\mathrm{tot}} = 2.55M_{\odot}$, the disc mass increases with aligned spins, reaching the maximum of $\sim0.4M_{\odot}$ for $M255_{\uparrow^{0.4} \uparrow^{0.4}}$ model, while decreasing with anti-aligned spins. However, for aligned spins beyond $\chi=0.4$, the disc mass begins to decrease but remains higher than the irrotational model. Single-spin and mixed spin models yield disc masses consistent with the general aligned  and anti-aligned models, with variations comparable to the numerical uncertainty between low and high resolution.
The relationship between the sum of individual spin magnitudes, $|\chi_{1}|+|\chi_{2}|$ and disc mass is also examined, in Fig.~\ref{fig:spindisc_fit_chisum}, as a potential method to constrain individual spins through electromagnetic counterparts. The results demonstrate that this relationship depends mildly on the total mass of the binary. For low mass BNS mergers $M_{\mathrm{tot}} = 2.55M_{\odot}$ that result in long-lived neutron stars, disc masses do not show a clear correlation with the sum of individual spin magnitudes. The observed flatness is not interpreted as a physical plateau, but rather as a weak or absent correlation for these models. 

\section{Results for models with $M_{\mathrm{tot}} = 3.05$ and $4.10\,M_{\odot}$}\label{results_other}
This section explores how the trends identified for different spins in the $M_{\mathrm{tot}}=2.55M_{\odot}$ models change when considering higher total masses and a different mass ratio.

\textbf{Inspiral trend: } The general inspiral trend persists for high mass models with anti-aligned (aligned) models merging earlier (later) than the irrotational model. For $M_{\mathrm{tot}}=3.05M_{\odot}$, the $M305_{\uparrow^{0.4} \uparrow^{0.4}}$ and $M305_{\uparrow^{0.67} \uparrow^{0.67}}$ models show the same change in trend with spin as for $M_{\mathrm{tot}}= 2.55 M_{\odot}$. The models with mass asymmetry merge earlier than the equal mass irrotational and aligned spin models with the same effective spin, due to lower total angular momentum and being tidally disrupted.

\textbf{Remnants: } The fate of the remnant is affected significantly by the total mass, mass asymmetry and spin. Among the models, only $M305_{\uparrow^{0.67} \uparrow^{0.67}}$ and $M305q205_{\uparrow^{0.6} {0}}$, result in long-lived neutron stars similarly to the $M_{\mathrm{tot}} = 2.55M_{\odot}$ models. On the other hand, the $M305_{\uparrow^{0.4} \uparrow^{0.4}}$ model results in a delayed collapse to a BH. All other models, regardless of spin, result in prompt BH formation. We test, albeit at low resolution, the upper limit of the black hole spin from BNS mergers. Our analysis shows that $M410_{\uparrow^{0.67} \uparrow^{0.67}}$ model achieves a dimensionless spin parameter of $\chi=0.92$, becoming the highest spinning BH produced by BNS mergers to date. This surpasses the previously reported limit of $\chi=0.888\pm0.018$ by~\citet{kastaun_black_2013}. Furthermore, our finding regarding the decrease in BH spin due to delayed collapse and mass asymmetry align with the results of~\citet{bernuzzi_how_2016} and~\citet{dietrich_massratio}, respectively. 
  
\textbf{Thermodynamic Properties: }For the high mass models, meaningful comparison with $M_{\mathrm{tot}}= 2.55M_{\odot}$ can only be made for models that form long-lived neutron stars. The $M305_{\uparrow^{0.67} \uparrow^{0.67}}$ model shows stronger shock heating and compression than $M255_{\uparrow^{0.67} \uparrow^{0.67}}$, reaching higher maximum temperature and density due to the increased total mass. The asymmetric mass model $M305q205_{\uparrow^{0.6} {0}}$ reaches a maximum density and temperature of $\sim 5.4\, \rho_{\mathrm{sat}}$ and $\sim78$ MeV, values which are significantly lower in temperature and only slightly lower in density than the $M255_{\downarrow^{0.65} \downarrow^{0.65}}$ model, despite the higher total mass of the asymmetric mass model. This highlights that anti-aligned spins enhance compression more strongly than aligned spins with mass asymmetry, even at higher total mass. Notably, while the $M255_{\downarrow^{0.65} \downarrow^{0.65}}$ model reaches its maximum density almost immediately after merger, the asymmetric mass model shows a delayed density peak, as material gradually accretes. This could be relevant for future EoS studies.

\textbf{Neutrinos:} Same as in the $M_{\mathrm{tot}}=2.55M_{\odot}$ models, the mean energy hierarchy is unaffected by total mass or mass asymmetry, consistently showing $\langle E_{\nu_x} \rangle > \langle E_{\bar{\nu}_e} \rangle > \langle E_{\nu_e} \rangle$. In contrast, the luminosity hierarchy varies with total mass and mass ratio, reflecting the thermodynamical evolution. For example, the $M305_{\uparrow^{0.67} \uparrow^{0.67}}$ model shows enhanced heavy-lepton production relative to electron neutrinos, yielding $L_{\bar{\nu}_e} > L_{\nu_x} > L_{\nu_e}$ instead of the $L_{\bar{\nu}_e} > L_{\nu_e} > L_{\nu_x}$ for $M255_{\uparrow^{0.67} \uparrow^{0.67}}$. The asymmetric mass model $M305q205_{\uparrow^{0.6} {0}}$, unlike the single-spin aligned model with $M_{\mathrm{tot}}=2.55M_{\odot}$, shifts to $L_{\nu_x} >L_{\bar{\nu}_e} > L_{\nu_e}$, resembling the pattern observed in irrotational and anti-aligned models with $M_{\mathrm{tot}}=2.55M_{\odot}$.

\textbf{Gravitational waves: } The trend that the irrotational model radiates the highest energy and angular momentum, as seen for $M_{\mathrm{tot}}=2.55M_{\odot}$, generally holds for higher mass models.
However, for $M_{\mathrm{tot}}=3.05M_{\odot}$, model $M305_{\uparrow^{0.4} \uparrow^{0.4}}$ with a delayed collapse, the radiation exceeds that of the irrotational model, which is consistent with additional post-merger emission preceeding BH formation (as calculated at $\sim10$ ms after collapse). Increasing the amount of aligned spin reduces the energy and angular momentum loss, as seen in the $M_{\mathrm{tot}}=2.55M_{\odot}$ models. Comparing $M305_{\uparrow^{0.4} \uparrow^{0.4}}$ and asymmetric mass model $M305q205_{\uparrow^{0.6} {0}}$ shows that mass asymmetry strongly suppresses energy and angular momentum losses, even given the same effective spin. For $M_{\mathrm{tot}}=4.10M_{\odot}$, all models promptly collapse to a BH, with the irrotational model showing the largest emission, same as for $M_{\mathrm{tot}}=2.55M_{\odot}$.

\textbf{Ejecta and disc: } For both $M_{\mathrm{tot}}=3.05$ and $4.10M_{\odot}$, models with anti-aligned spin promptly collapse to a BH and produce a negligible amount of ejecta or none at all. For $M_{\mathrm{tot}}=3.05M_{\odot}$, mass asymmetry enhances the total amount of ejecta and disc mass through tidal disruption, even in irrotational models. For $M_{\mathrm{tot}}=2.55M_{\odot}$ models, an increase of spin from $\chi=0.4$ to $0.67$ results in an increase of the total ejecta amount, but reduces the disc mass, whereas for $M_{\mathrm{tot}}=3.05M_{\odot}$ the disc mass continues to increase, reaching $\sim0.26M_{\odot}$. In aligned spin models, fast-moving ejecta is absent for $M_{\mathrm{tot}}=2.55$ and $3.05M_{\odot}$ and negligible for $M_{\mathrm{tot}}=4.10M_{\odot}$. Overall, disc mass decreases with increasing total mass. This is likely due to prompt collapse reducing the available material. More importantly, even in models that undergo prompt collapse to a BH, mass asymmetry and aligned spins contribute to increased amount of ejecta and disc mass, consistent with~\citet{dietrich_massratio}, who considered irrotational models.

\begin{figure}
\centering
\includegraphics[width=1.0\columnwidth]{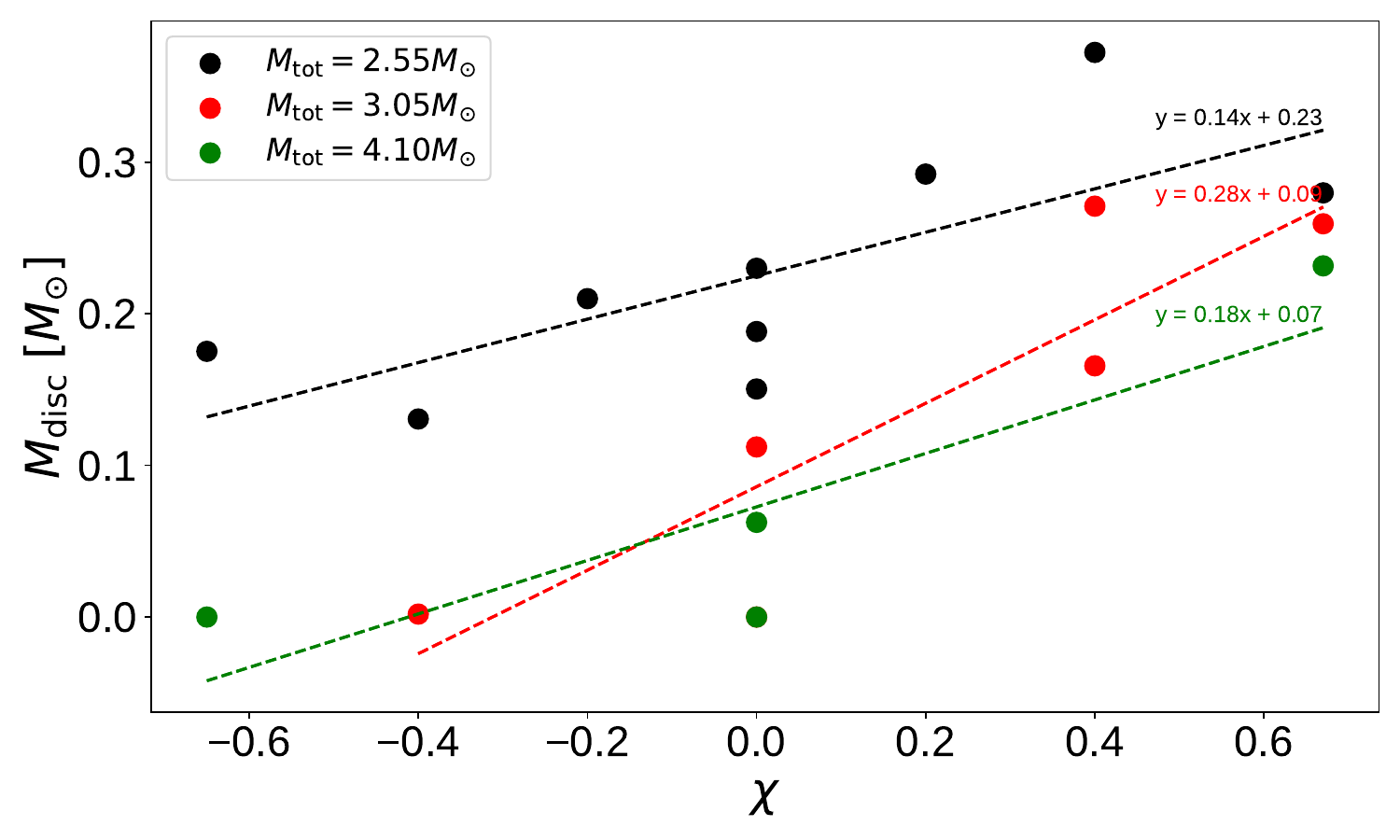}
\caption{
The relationship between effective spin and disc mass. Circles represent different total mass models, $M_\mathrm{tot}=2.55M_\odot$ (black), $M_\mathrm{tot}=3.05M_\odot$ (red) and $M_\mathrm{tot}=4.10M_\odot$ (green). The dashed lines indicate linear regression fits, showing the dependence of disc mass on effective spin and total mass. Even though we have six simulations for $M_\mathrm{tot}=3.55 M_{\odot}$, one data point appears to be missing due to the overlap at $\chi = 0$ between two models ($M305_{{0} {0}}$ and $M410_{{0} {0}}$). These fits are included solely to illustrate general trends.
        }
\label{fig:spindisc_fit}
\end{figure}

\begin{figure}
    \centering
    \includegraphics[width=1.0\columnwidth]{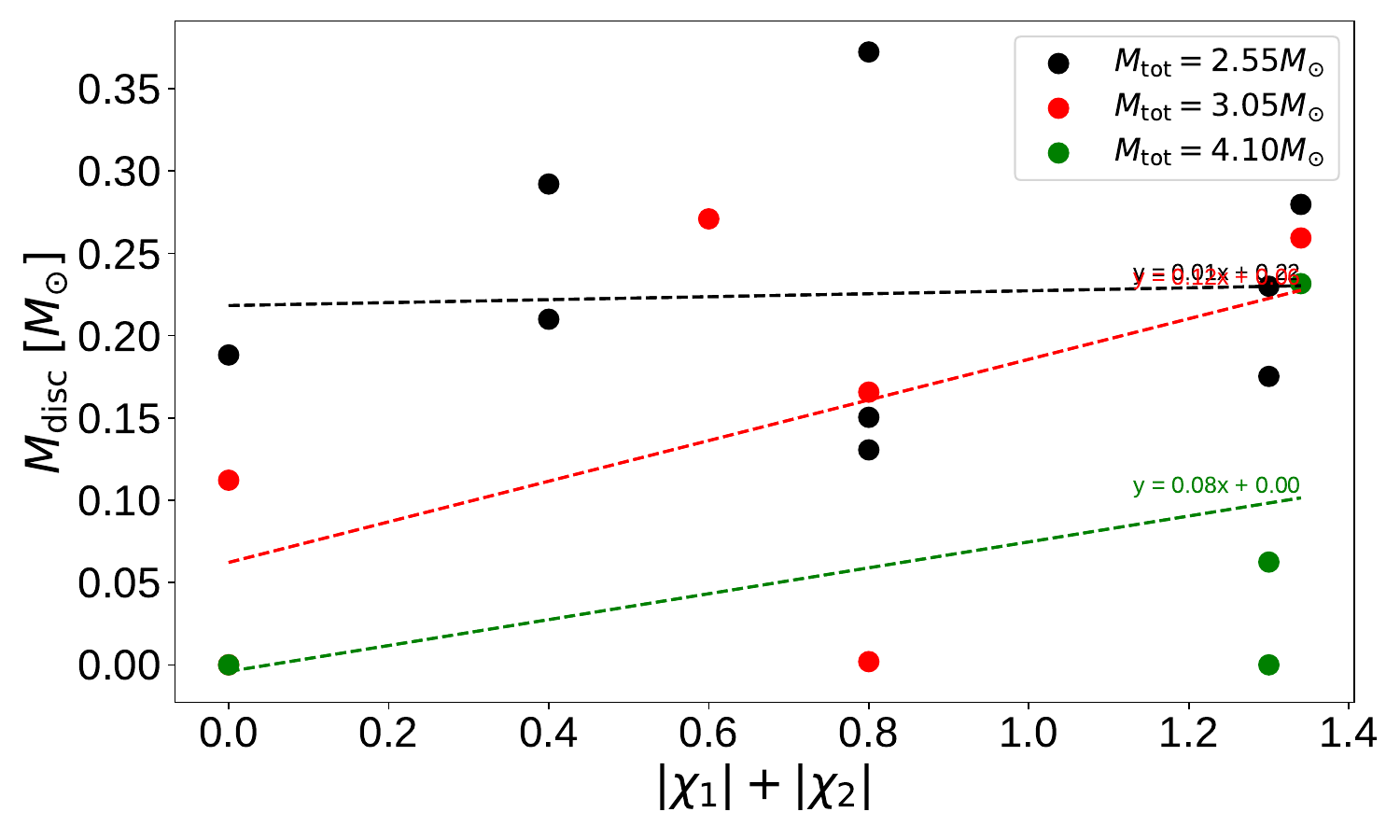}
    \caption{
    Similar to Fig.~\ref{fig:spindisc_fit}, but showing the relationship between the sum of individual spin magnitudes and disc mass. The dashed lines represent linear regression fits, included to visualize possible trends. As in Fig.~\ref{fig:spindisc_fit}, there is an apparent overlap at $|\chi_{1}|+|\chi_{2}| = 0 $ between models $M305_{{0} {0}}$ and $M410_{{0} {0}}$, both of which yield very low disc masses. These fits are included solely to illustrate general trends.
        }
        
    \label{fig:spindisc_fit_chisum}
\end{figure}

For $\chi=0.67$, the amount of total ejecta increases slightly comparing $M_{\mathrm{tot}}=2.55$ to $3.05M_{\odot}$, but decreases again for $M_{\mathrm{tot}}=4.10M_{\odot}$ models, where $M410_{\uparrow^{0.67} \uparrow^{0.67}}$ still yields the largest ejecta and disc mass for the models with the same total mass. This shows that while a high total mass generally suppresses ejecta due to prompt collapse, such high and aligned spin models can still produce a massive amount of ejecta. We note that there is an apparent overlap between models $M305_{{0} {0}}$ and $M410_{{0} {0}}$ in Fig.~\ref{fig:spindisc_fit} and Fig.~\ref{fig:spindisc_fit_chisum}, as both models yield very low disc masses (on the order of $10^{-5} M_{\odot}$). See Table~\ref{ejecta_properties_table} for the exact values.

We consider the impact of different total masses and mass asymmetry on the relationship between the sum of individual spin magnitudes, $|\chi_{1}|+|\chi_{2}|$, and disc mass, as shown in Fig.~\ref{fig:spindisc_fit_chisum}. For low mass models ($M_{\mathrm{tot}}=2.55M_{\odot}$), the disc mass remains nearly constant across different spin magnitudes. In contrast, for higher mass models ($M_{\mathrm{tot}}=3.05M_{\odot}$ and $4.10M_{\odot}$), the disc mass exhibits a slight increasing trend with spin magnitude. These findings suggest that electromagnetic counterparts could offer valuable insights into the individual spins of neutron stars by measuring disc mass, particularly models undergoing prompt collapse to a black hole in equal mass, irrotational configurations.

\section{Discussion and Conclusion}\label{conclusion}
    
We study the effect of the spin in BNS mergers. We consider $19$ configurations with three different total masses and having both equal and unequal mass. Also spin configurations are varied, to cover cases where both stars' spins are aligned, or anti-aligned or mixed, with respect to the orbital spin. We investigate the impact of the initial spin on gravitational radiation emission, on properties of ejected matter and on thermodynamic properties of all these systems as well as gravitational wave strains, their detectability, the maximum temperature, the rest-mass density of matter, the neutrino energies, luminosities and disc masses.

Spin fundamentally alters the orbital dynamics, neutron star structure, gravitational wave emission, ejecta, and disc masses. Its magnitude and orientation influences the duration of the inspiral phase through spin-orbit, spin-spin, and tidal interactions. We find that spin-orbit interactions dominate until $|\chi| = 0.4$, with aligned (anti-aligned) spins extend (shorten) the inspiral phase due to their attractive (repulsive) behaviour, a trend consistent with previous studies~\citep{kastaun_black_2013, tsatsin_initial_2013, bernuzzi_mergers_2014, Dietrich_2017_rotation, East_2019}. Beyond this threshold, spin-spin interactions become significant and counteract the effects of spin-orbit interactions, leading to earlier (later) mergers for aligned (anti-aligned) spins. Although this change in trend is observed in both low and high resolutions, its significance may diminish at higher resolutions.

Thermodynamic properties, such as maximum rest-mass density and temperature, are unaffected by this change in trend. Aligned (anti-aligned) spins consistently result in less (more) violent mergers, reaching lower (higher) maximum rest-mass densities and temperatures compared to the irrotational model. These values peak at $\sim145$ MeV and $5.6 \rho_{\mathrm{sat}}$ for the highly anti-aligned spin model. Neither the spin magnitude nor its orientation alters the energy hierarchy among neutrino flavours: heavy-lepton neutrinos are the most energetic, while electron neutrinos are the least energetic. Total neutrino energies and luminosities do not show a monotonic trend with spin orientation. Aligned (anti-aligned) spins tend to suppress (enhance) the overall neutrino number flux, resulting in lower (higher) neutrino energies and luminosities. This behavior also alters the luminosity hierarchy among flavours, with aligned spins favouring dominant $\bar{\nu}_e$ emission, while anti-aligned spins preserve the typical hierarchy with $\nu_x$ remaining dominant.

Spin significantly influences the structure of the remnant neutron star, with aligned spin producing extended spiral arms as a result of redistribution of additional angular momentum. Anti-aligned spins, on the other hand, lead to a significant elongation of the stars prior to merger, consistent with~\citet{Dudi_Dietrich2022}, who studied the maximum anti-aligned spin of $\chi=-0.28$. For aligned spins, the additional angular momentum increases the rotational support of the remnant, in agreement with~\citet{East_2019, Chaurasia_Dietrich2020}, while anti-aligned spins reduce rotational support, consistent with~\citet{Dudi_Dietrich2022}. 

Spin asymmetries, such as models where only one component has spin or models with mixed aligned and anti-aligned spins, exhibit behaviour similar to mass asymmetry. This finding aligns with~\citet{rosswog_diener_frontiers, rosswog_2024}. Additionally, we observe that spin mimics the behaviour of different EoS; this agrees with~\citet{East_2019}, where degeneracies between spin and EoS were reported. Similarly,~\citet{Dietrich_2017_rotation} identified degeneracies between spin effect, mass ratios and EoSs, reporting that the influence of spin was smaller than the mass ratios considering spins of $\chi=0.1$.

Mergers with aligned spins radiate more energy and angular momentum through gravitational waves than anti-aligned spins, yet the irrotational model exhibits the highest overall energy and angular momentum release. We observe changes in both the one-arm and fundamental mode frequencies with spin. For models with $M_{\mathrm{tot}} = 2.55M_{\odot}$, fundamental mode frequencies shift to higher values for aligned spins compared to the irrotational model, consistent with~\cite{Dietrich_2017_rotation}, but contrasting with~\citet{bernuzzi_mergers_2014, East_2019, rosswog_2024}. The impact of increasing aligned spin on the frequency shift is minimal, around $\sim10$ Hz for $M_{\mathrm{tot}} = 2.55 M_{\odot}$, but increases to $\sim 400$ Hz for higher total mass models, highlighting the dependence of the shift on total mass. Conversely, fundamental mode frequencies shift to lower values for anti-aligned spins compared to the irrotational model. This behaviour differs from the findings by~\citet{East_2019}, who reported that aligned spins shift to lower frequencies and anti-aligned spins shift to higher frequencies for maximum spins of $-0.13$ (anti-aligned) and $0.33$ (aligned), as compared to the irrotational model. For the one-arm mode, our high-resolution simulations show that its frequency increases by at most $\sim30$ Hz for aligned spins, while it increases significantly up to $\sim1120$ Hz for the highly anti-aligned spin model.

Examining the rotation profile of the remnant neutron stars, we observe that the core rotates more slowly than the envelope, consistent with the findings of~\citet{shibata2006, kastaun_properties_2015, East_2019}. The maximum spin is reached for the irrotational model, also in agreement with~\citet{kastaun_properties_2015}. Additionally, we test the formation of the fastest spinning BH from BNS mergers and find that our $M410_{\uparrow^{0.67} \uparrow^{0.67}}$ model produces the fastest spinning BH to date, with a dimensionless spin of $\chi=0.92$, surpassing the previously reported limit of $\chi=0.888\pm0.018$ by~\citet{kastaun_black_2013}. However, it is important to emphasize that this result requires further confirmation with higher resolution.

Beyond its influence on remnant structure, spin significantly affects the total mass and composition of the ejecta, including its fast-moving component. The ejecta mass is strongly dependent on both the magnitude and orientation of the spin. All spin configurations studied result in higher total ejected mass compared to the irrotational model. The composition generally is more neutron rich for aligned spins than anti-aligned spins. Notably, we observe that if only one component has spin, the behaviour differs from models where both components are spinning. Specifically, in models with one spinning component, anti-aligned spin produces more ejecta than aligned spin, whereas in models where both components are spinning, aligned spins result in more ejected mass than anti-aligned models. In the literature, ~\cite{kastaun_properties_2015} reported that spinning models result in lower ejecta mass compared to the irrotational model, which contrasts with our findings, where the irrotational model yields the least total ejecta mass.~\citet{East_2019} found that anti-aligned spins result in more ejecta than aligned spins for spin in the range $\chi = -0.13$ to $0.33$, and similar results were reported by~\citet{Chaurasia_Dietrich2020} for $\chi = 0.096$. These studies are inconsistent with our findings. However, we do observe the presence of fast-moving ejecta in anti-aligned spins, which aligns with~\citet{East_2019}. Furthermore,~\citet{Dudi_Dietrich2022} who explored spins in the interval $\chi=-0.28$ to $0.58$, reported that aligned spins result in higher ejecta mass than anti-aligned spins, consistent with our results.

Spin also influences the disc mass. For $M_{\mathrm{tot}} = 2.55 M_{\odot}$, aligned spins lead to higher disc masses, while anti-aligned spins result in lower disc masses compared to the irrotational model. The increase in disc mass with aligned spins aligns with findings by~\citet{East_2019, Chaurasia_Dietrich2020, rosswog_2024, federico_dietrich_prompt}. Notably, the disc mass peaks at $\chi = 0.4$ for aligned spins and decreases beyond this value, whereas it continues to increase for anti-aligned spins. In contrast, for $M_{\mathrm{tot}}=3.05M_{\odot}$, the disc mass continues to increase even for aligned spins exceeding $\chi=0.4$, highlighting the influence of total mass on this relationship. This suggests that the effective spin could also be constrained using electromagnetic counterparts, providing an additional avenue to complement gravitational wave observations.

In addition to the dependence of disc mass on effective spin, as shown in Fig.~\ref{fig:spindisc_fit}, where disc mass increases (decreases) with positive (negative) effective spin, we analyse its variation with the sum of individual spin magnitudes, $|\chi_{1}|+|\chi_{2}|$, in Fig.~\ref{fig:spindisc_fit_chisum} as a potential method to constrain individual spins. We find that this relationship depends on the total mass of the binary. For $M_{\mathrm{tot}}=2.55M_{\odot}$, the disc mass remains relatively constant across different spins, while higher total mass models exhibit an increase in disc mass with spin. These findings emphasize the importance of considering high spin configurations when investigating electromagnetic counterparts, even for equal mass binaries that undergo prompt collapse to a BH. Since disc mass is the quantity with the most impact on the kilonovae peak luminosity, this finding suggests that one could use EM data to constrain $|\chi_{1}|+|\chi_{2}|$, breaking the degeneracy in the measurement of spin from gravitational wave alone for models with total mass that leads to prompt BH formation for equal mass, irrotational model \footnote{David Radice, personal communication}.

Although the spin values and mass ratio investigated in this study are higher than what has been observed, we demonstrate the significant impact of such high spins on gravitational wave emissions and properties of ejected matter. These findings suggest that systems with high spins can be identified through gravitational wave observations. However, the differences in potential electromagnetic counterparts are not explored in this study.
  
\section*{Acknowledgements}
\addcontentsline{toc}{section}{Acknowledgements}
We would like to thank the anonymous referee for their time provided generously, their invaluable comments and multiple very detailed suggestions all of which significantly improved both the content and the readability of this paper.
RM and BK would like to thank Ian Hawke for his comments \& suggestions, guidance and support. BK would like to thank Roland Haas for his constant support, guidance and his useful comments \& suggestions on every part of the study, David Radice for his many useful comments \& suggestions, help and guidance, Wolfgang Tichy for his support and guidance, and Zachariah B.~Etienne for his valuable comments.
    
Part of the study was presented at Thematic school GWsNS-2023: Gravitational Waves from Neutron Stars and at the European Einstein Toolkit Meeting 2023.
    
This work used TACC Stampede at XSEDE through allocation PHY160053 from the Extreme Science and Engineering Discovery Environment (XSEDE), which was supported by National Science Foundation grant number \#1548562. The work has been performed under the Project HPC-EUROPA3  (INFRAIA-2016-1-730897), with the support of the EC Research Innovation Action under the H2020 Programme; in particular, BK gratefully acknowledges the support of The University of Edinburgh and the computer resources and technical support provided by EPCC. The numerical calculations reported in this paper were partially performed at TUBITAK ULAKBIM, High Performance and Grid Computing Center (TRUBA resources).
    
\section*{Data Availability}
    
The data generated in this article is available for a reasonable request from the corresponding author.
    
\bibliographystyle{mnras}
\bibliography{dumbledore} 

\end{document}